\documentclass[floatfix,%
 aip,
 amsmath,amssymb,
 reprint,%
]{revtex4-1}
\usepackage{graphicx}% Include figure files
\usepackage{dcolumn}% Align table columns on decimal point
\usepackage{bm}% bold math
\usepackage[utf8]{inputenc}
\usepackage[T1]{fontenc}
\usepackage{mathptmx}
\usepackage{etoolbox}

\usepackage[compatibility=false]{caption}
\usepackage{subcaption}
\usepackage{newunicodechar}
\newunicodechar{−}{\ensuremath{-}}

\usepackage{amsmath} % Added for bmatrix environment

\usepackage{subcaption}
\usepackage[normalem]{ulem}

\makeatletter
\def\@email#1#2{%
 \endgroup
 \patchcmd{\titleblock@produce}
  {\frontmatter@RRAPformat}
  {\frontmatter@RRAPformat{\produce@RRAP{*#1\href{mailto:#2}{#2}}}\frontmatter@RRAPformat}
  {}{}
}%
\makeatother

\usepackage{hyperref}
\begin{document}
\preprint{AIP/123-QED}

\title[Quantum Radar: An Engineering Perspective]{Quantum Radar: An Engineering Perspective\texorpdfstring{\\}{ }}
% Force line breaks with \\
\author{Murat Can KARAKOÇ}
\email{murat.karakoc@erzurum.edu.tr}
\affiliation{Department of Electrical and Electronics Engineering, Erzurum Technical University, Erzurum, Türkiye}

\author{Özgün ERSOY}
\affiliation{Department of Electrical and Electronics Engineering, Ankara Yıldırım Beyazıt University, Ankara, Türkiye}

\author{Ahmad SALMANOGHLI KHIAVI}
\affiliation{Department of Electrical and Electronics Engineering, Ankara Yıldırım Beyazıt University, Ankara, Türkiye}

\author{Asaf Behzat ŞAHİN}
\affiliation{Department of Electrical and Electronics Engineering, Ankara Yıldırım Beyazıt University, Ankara, Türkiye}

\date{\today}% It is always \today, today,

\begin{abstract}
Quantum radar has emerged as a promising paradigm that utilizes entanglement and quantum correlations to overcome the limitations of classical detection in noisy and lossy environments. By exploiting microwave entanglement generated from superconducting devices such as Josephson parametric amplifiers, converters, and traveling-wave parametric amplifiers, quantum radar systems can achieve enhanced detection sensitivity, lower error probabilities, and greater robustness against thermal noise and jamming. This review provides a comprehensive overview of the field, beginning with the theoretical foundations of quantum illumination and extending to the generation of entanglement in the microwave regime. We then examine key quantum radar subsystems, including quantum transducers, amplification chains, and receiver architectures, which form the backbone of practical designs. Recent experimental systems are surveyed in the microwave domain, highlighting proof-of-principle demonstrations and their transition from conceptual frameworks to laboratory realizations. Collectively, the progress reviewed here demonstrates that quantum radar is evolving from a theoretical construct to a practical quantum technology capable of extending the performance boundaries of classical radar.
\end{abstract}

\maketitle

%\begin{quotation}
%The ``lead paragraph'' is encapsulated with the \LaTeX\ 
%\verb+quotation+ environment and is formatted as a single paragraph before the first section heading. 
%(The \verb+quotation+ environment reverts to its usual meaning after the first sectioning command.) 
%Note that numbered references are allowed in the lead paragraph.
%
%The lead paragraph will only be found in an article being prepared for the journal \textit{Chaos}.
%\end{quotation}

\section{Introduction} \label{sec1}

Quantum radar systems propose to exceed the standard classical limit in terms of sensitivity and resolution through the utilization of quantum resources such as entanglement and squeezing \cite{lloyd2008enhanced}. This advantage is especially relevant for detecting weakly reflecting targets under strong thermal or electronic noise, where classical systems struggle. 

The primary difference lies in the nature of the transmitted and received signals. Classical radar transmits coherent microwave pulses and detects echoes through direct energy return and correlation with the transmitted waveform \cite{skolnik1980introduction}. Quantum radar, on the other hand, uses entangled photon pairs, where the signal is transmitted toward the target and the idler is stored \cite{lloyd2008enhanced, tan2008quantum, shapiro2020quantum}. Detection is not based on the direct return of the signal but on correlations between the returned signal and the retained idler, enabling discrimination of the target even under severe noise or loss conditions \cite{guha2009gaussian}.

In optical domain implementations of quantum radar, entangled photon pairs are typically generated through spontaneous parametric down-conversion (SPDC) or four-wave mixing (FWM). The signal photon is directed toward a target, while the idler (or herald) photon is retained for correlation analysis. Notable demonstrations include standoff target detection using correlated photon pairs with timing-coincidence measurements, even under strong environmental noise \cite{lopaeva2013experimental, zhang2015entanglement, england2019quantum, liu2019enhancing, zhao2022light}.

Various methods have been proposed to quantify the entanglement of quantum states in radar systems. Commonly used approaches include the symplectic eigenvalue test \cite{serafini2007detecting}, the Peres – Horodecki – Simon criterion \cite{simon2000peres}, and measures of quantum discord \cite{ollivier2001quantum, henderson2001classical}, each providing information on how quantum correlations behave under noise and loss.

The concept of microwave quantum illumination \cite{barzanjeh2015microwave} is based on foundational studies in reversible optical-to-microwave quantum interfaces \cite{barzanjeh2012reversible} and strong coupling in circuit cavity electromechanics \cite{teufel2011circuit}, yet it remains a theoretical proposal without experimental realization. Nevertheless, it represents a paradigm shift in quantum radar, transitioning the operational domain from optics to the microwave regime, where classical radar systems operate but quantum advantages had not been practically demonstrated. Subsequent proposals introduced optoelectronic converters as an alternative route to maintain entanglement at higher temperatures and reduce reliance on mechanical interfaces \cite{salmanogli2019entanglement}. Some experimental demonstrations using Josephson parametric devices, such as Josephson parametric amplifier (JPA), Josephson parametric converter (JPC), and Josephson traveling-wave parametric amplifier (JTWPA), have transformed quantum illumination from a theoretical proposal into a viable microwave sensing strategy \cite{chang2019quantum, barzanjeh2020microwave, livreri2022microwave}.

The atmospheric and channel effects critically influence the performance of quantum radar. Although microwave frequencies experience less attenuation and scattering than optical frequencies, thermal noise, absorption, and turbulence can still degrade entanglement and weaken correlations \cite{hoijer2019quantum, salmanogli2019modification, salmanogli2020analysis}. These processes are often modeled with cascaded beam splitters, showing how thermally excited photons reduce fidelity and shorten the entanglement lifetimes \cite{jeffers1993quantum, salmanogli2020optoelectronic}. Despite these challenges, recent work shows that entanglement-based sensing protocols can retain advantages over classical systems under realistic noise and loss conditions \cite{zhang2013entanglement, zhang2015entanglement}.

Since ideal joint measurements remain technologically challenging, experimental systems rely on hybrid receivers in which signal and idler are downconverted, digitized, and processed using high-speed analog-to-digital converters (ADCs) and FPGA-based platforms \cite{luong2019receiver, barzanjeh2020microwave}. This digital architecture enables real-time quadrature extraction, second-order moment analysis, and flexible post-processing strategies that approximate the performance of optimal quantum receivers. 

In the following sections of this paper are organized as follows. In Sect.  \ref{sec2}, we summarize the fundamental principles of quantum illumination and its distinction from classical detection. We then discuss entanglement generation in the microwave regime, focusing on Josephson-based devices and other techniques for producing correlated microwave photons. Sect. \ref{sec3} highlights critical building blocks such as quantum transducers, parametric amplifiers, atmospheric effects, path loss and receiver architectures. Next, in Sect. \ref{sec4} surveys existing implementations and proof-of-principle demonstrations, illustrating how theoretical concepts are implemented in practice. Finally, the paper concludes with a in Sect. \ref{sec5} that outlines the current challenges, potential applications, and future directions, followed by an abbreviation list and references.

\section{Theoretical Backgrounds}  \label{sec2}

\subsection{Quantum Correlation and Non-Classicality As An Essential Key in Quantum Radar}

Classical radar detects objects by comparing the amplitude and phase of reflected electromagnetic waves against a reference signal. The performance of such systems is ultimately constrained by classical noise statistics. Quantum radar, in contrast, attempts to take advantage of non-classical correlations, particularly quantum entanglement, to obtain lower error-probability in target detection, even when entanglement is largely destroyed during propagation. Entanglement turns a pair of photons into a single, delocalized information carrier, so that information acquired by measuring the return signal can be interpreted only in the light of the locally stored idler.  This global description underpins protocols such as quantum illumination (QI), which theoretically achieves a signal-to-noise (SNR) advantage over any classical scheme that irradiates the same mean number of photons \cite{lloyd2008enhanced, barzanjeh2015microwave}. Entanglement was introduced in the 1935 Einstein–Podolsky–Rosen (EPR) paper as a challenge to the completeness of quantum mechanics \cite{einstein1935can}. Schrödinger coined the term “entanglement” to describe states that cannot be expressed as a product of local states for individual subsystems \cite{schrodinger1935discussion}. A pure bipartite state $|\Psi\rangle$ is entangled when its Schmidt number exceeds one \cite{nielsen2010quantum}. For mixed states, separability is defined by the criterion $\rho=\Sigma_{\mathrm{i}} \mathrm{p}_{\mathrm{i}} \rho_{\mathrm{Ai}} \otimes \rho_{\mathrm{Bi}}$; any state that cannot be decomposed in this form is entangled. Bell’s 1964 theorem confirmed that the measurement statistics of the entangled states can violate the inequalities obeyed by all local‑realist theories, thereby revealing intrinsically non‑classical correlations \cite{bell1964einstein}. In the radar context, these correlations are exploited not for provable non‑locality but for enhanced hypothesis testing in noisy channels. In the following sections, various versatile criteria used to quantify quantum correlations will be discussed and analyzed in detail.

\subsubsection{Selected Criteria for Quantum Correlation in Quantum Radar Systems}

It should be noted that through the literature survey, it is found that there are a lot of versatile methods and approaches to study non-classicality or quantum correlation in a quantum system; some bold utilized criteria are the symplectic eigenvalue ($2\eta$) \cite{salmanogli2019entanglement, tasgin2020measuring}, Peres-Horodesky-Simon \cite{tasgin2020measuring, ge2015conservation, simon2000peres}, and quantum discord \cite{salmanogli2022entangled, salmanogli2023enhancing, barzanjeh2020microwave}; however, the reader may find other reported criteria as well. Next, we briefly discuss these criteria and attempt to apply some of them to the quantum system used in quantum radar applications.

\paragraph{Peres-Horodecki-Simon Criterion ($\lambda_S$):} The Peres-Horodecki-Simon criterion is a foundational separability test used in the study of quantum entanglement, especially in CV systems \cite{simon2000peres}. It originates from the Peres criterion \cite{peres1996separability}, which states that a necessary condition for separability is that the partial transpose of the density matrix of the system remains positive (positive partial transpose, PPT). While originally framed for finite-dimensional systems, Simon extended this concept to Gaussian states using covariance matrices \cite{simon2000peres}. In this context, the criterion involves applying partial transposition to the covariance matrix and checking $\lambda_S > 1$ for entanglement. This criterion is the necessary and sufficient condition for separability of the bipartite Gaussian states. Thus, it should be assumed that all the states of the cavity are Gaussian. The Peres-Horodecki-Simon test is both practical and computationally efficient, making it a standard tool for analyzing CV entanglement in quantum optics, cavity QED, and other Gaussian state-based quantum systems \cite{tasgin2020measuring, ge2015conservation, simon2000peres, peres1996separability, horodecki1996separability, duan2000inseparability}.

\paragraph{Quantum Discord:} Quantum discord is a measure of quantum correlations in a bipartite quantum system that go beyond entanglement. It captures the difference between two classically equivalent expressions for mutual information when extended to the quantum domain \cite{ollivier2001quantum, henderson2001classical, giorda2010gaussian}. Although classical systems have only one definition of mutual information, quantum systems can exhibit discrepancies due to measurement-induced disturbance. Discord quantifies the minimum loss of correlations after a local measurement in a subsystem. A nonzero discord implies the presence of non-classical correlations, even in separable (nonentangled) states, making it a more general indicator of quantumness than entanglement alone.

\begin{figure}
    \centering
    \includegraphics[width=0.25\textwidth]{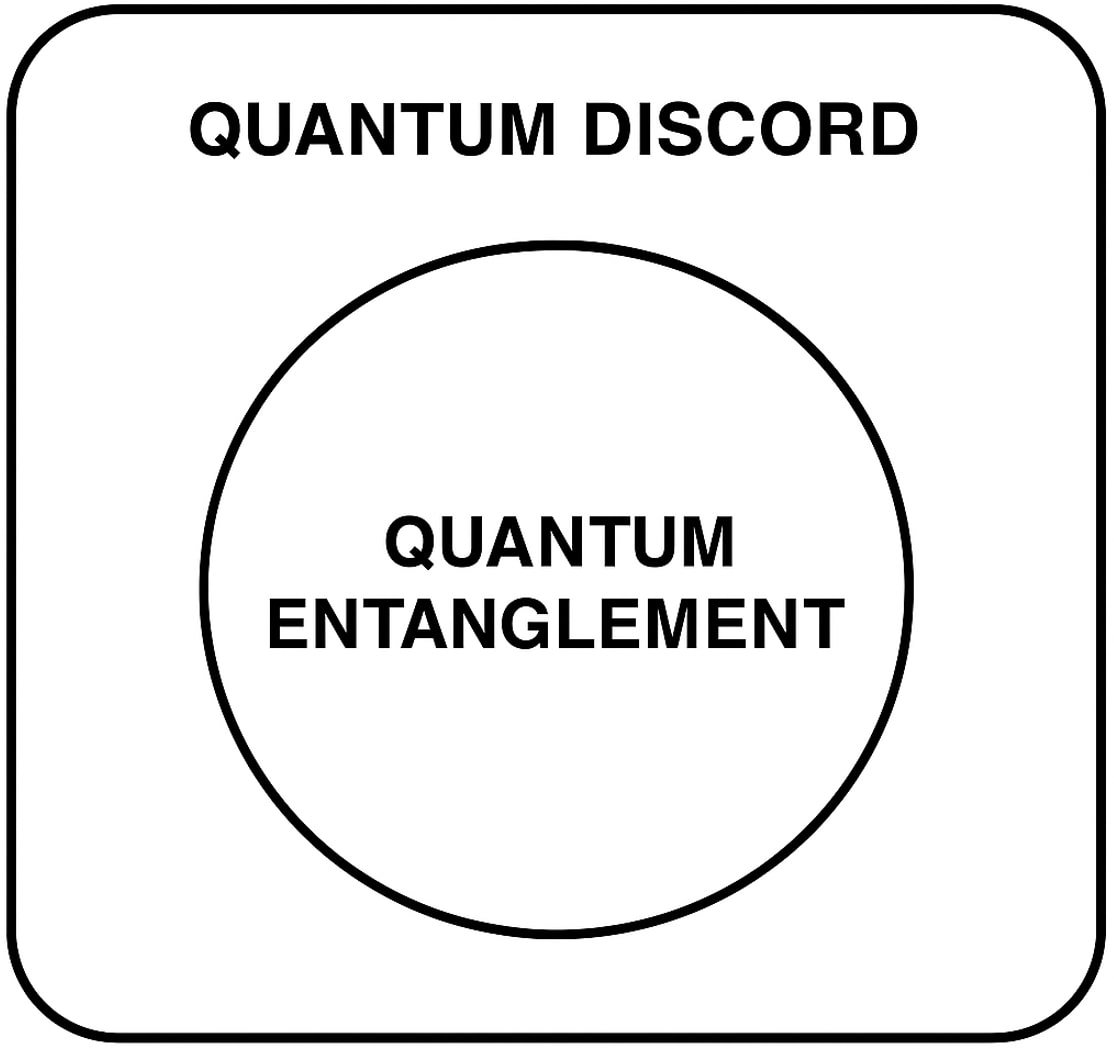}
    \caption{\label{fig:q_discord} Diagram representing the relationship between quantum entanglement and quantum discord. All entangled states lie within the set of discordant states, but discord can exist without entanglement.}
\end{figure}

Although both quantum discord and entanglement measure non-classical correlations, they differ fundamentally. Entanglement is a strict subset of quantum correlations and requires inseparability of the quantum state. Quantum discord, however, can be nonzero even in separable (unentangled) mixed states. This means that systems that are not entangled may still exhibit quantum behavior via discord. Entanglement is typically destroyed by decoherence, whereas discord is more robust in noisy environments. Thus, discord encompasses a broader class of quantum correlations, particularly useful when entanglement is too fragile or absent. In practical terms, discord can exist in situations where entanglement-based strategies fail \cite{ollivier2001quantum, henderson2001classical, giorda2010gaussian}.For comparison, one can consider Fig. \ref{fig:q_discord}. Although entanglement enables high-precision sensing and non-classical signal advantages, it is often fragile under loss and noise, common in realistic radar channels. Quantum discord, being more resilient, offers a compelling alternative. Even when entanglement is lost, the discord may persist and contribute to enhanced target detection or reduced error probabilities. Therefore, discord-based protocols are potentially more robust for practical quantum radar applications. They provide a way to exploit quantum advantages without relying solely on entanglement, making them suitable for noisy open-system scenarios typical of radar operations \cite{barzanjeh2015microwave, barzanjeh2020microwave}.

The calculation of quantum discord begins using two generalizations of classical mutual information \cite{ollivier2001quantum, henderson2001classical, giorda2010gaussian, adesso2010quantum}. Mutual information is used primarily to evaluate the total correlations between the subsystems. In the first generalization, the quantum mutual information for two systems, $A$ and $B$, is defined as $I(\rho_{AB}) = S(\rho_A) + S(\rho_B) - S(\rho_{AB})$, where $S(\rho_A) = -Tr(\rho_Alog_2\rho_A)$ is the von Neumann entropy of system $A$, and $I(\rho_{AB})$ is the conditional von Neumann entropy \cite{ollivier2001quantum, henderson2001classical, giorda2010gaussian, adesso2010quantum}. In the following, for the system studied in this work, the first oscillator is called in a short way $A$, and the second is called $B$. Conditional entropy arises because the measurement process disturbs the state in which a physical system is set. In other words, the measurement applied to the subsystem $B$ can change the state of the subsystem $A$.  

The second generalization introduces the entropic quantity $C(A|B)$, by which the classical correlation is calculated in the joint state $\rho_{AB}$. The classical correlation defines the maximum information about one subsystem depending on the measurement types applied to the other subsystem. The entropic quantity, considering the generalization, is defined as $C(\rho_{AB}) = S(\rho_A) - S_{min(\rho_{AB})}$, where the only difference between mutual quantum information is $S_{min}(\rho_{AB})$. That is generally described as positive operator-valued measures (POVMs) \cite{ollivier2001quantum, henderson2001classical, giorda2010gaussian, adesso2010quantum}. Thus, quantum discord is usually defined as $D(\rho_{AB}) = I(\rho_{AB}) - C(\rho_{AB})$, and substituting from the above definitions it becomes $D(\rho_{AB}) = S(\rho_B) - S(\rho_{AB}) + S_{min}(A|B)$. 

\subsection{Entanglement Generation in the Microwave Regime}

\subsubsection{Electro-Opto-Mechanical Converter : Simon-Peres-Horodecki Criterion}

 In this section, we reviewed various methods that have been employed in quantum radar systems to generate entangled microwave photons, and explored different approaches to analyze the resulting non-classicality. Since generating entanglement directly in the microwave regime is experimentally challenging, one promising method is to first generate optical entanglement and then convert part of the entangled state into the microwave domain. This necessitates a hybrid quantum interface capable of mediating entanglement between optical and microwave photons. In this direction, we focus on an electro-opto-mechanical (EOM) converter \cite{andrews2014bidirectional, safavi2011proposal, aspelmeyer2014cavity}, a system that couples optical and microwave cavities via a common mechanical resonator. In the EOM system, the mechanical element—typically a vibrational mode of a nanomembrane or drumhead—acts as a quantum bus, enabling coherent transfer of energy and quantum correlations between the optical and microwave domains. This allows us to capture both the coherent interactions and the noise dynamics critical for evaluating the entanglement transfer efficiency. Such a system plays a crucial role in future quantum networks and sensing platforms, where coherent microwave-optical interfaces are essential. The performance of EOM converters can be quantified by metrics such as cooperativity, entanglement generation rate, and transfer fidelity.

\begin{figure} [htbp]
    \centering
    \includegraphics[width=0.45\textwidth]{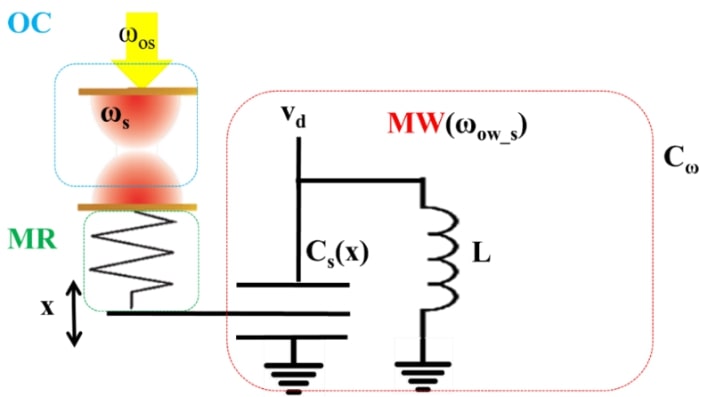}
    \caption{\label{fig:q_eom} EOM converter; coupling subsystems contain OC, MC, and MR \cite{salmanogli2020entanglement}.}
\end{figure}

To analyze the dynamics of the system, we employ the quantum electrodynamics (QED) framework as outlined in the foundational studies \cite{huttner1992quantization,salmanogli2020optical}. The analysis begins by formulating the equations of motion for the EOM converter, which constitutes a key subsystem of a typical quantum radar architecture, schematically depicted in Fig. \ref{fig:q_eom}. Following standard QED methodology, the Lagrangian formalism is used to systematically describe the system's behavior. The converter in Fig. \ref{fig:q_eom} comprises three core components: the optical cavity (OC), the mechanical resonator (MR), and the microwave cavity (MC), all of which are dynamically coupled. To rigorously capture the full dynamics, we must first construct individual Lagrangians for each of these subsystems and then account for the interaction Lagrangians that describe their mutual couplings. This step is critical because it allows us to derive the full quantum-mechanical description of the system, including entanglement and energy exchange processes. Accordingly, the total Lagrangian of the EOM system is composed as follows:

\begin{equation}
    \begin{array}{l}
        L_{O C}=\frac{\varepsilon_{0}}{2}\left(\dot{A}^{2}-\omega_{c}^{2} A^{2}\right) \\[6pt]
        L_{M R}=\frac{1}{2} m \dot{X}^{2}-\frac{1}{2} m \omega_{m}^{2} X^{2} \\[6pt]
        L_{M C}=\frac{1}{2} C(x) \dot{\phi}^{2}-\frac{1}{2 L} \phi^{2}+\frac{1}{2} C_{d}\left(v_{d}-\dot{\phi}\right)^{2} \\[6pt]
        L_{O C-M R}=-\alpha_{c} A \dot{X}
    \end{array}
    \label{eq:cavity_Lagrangian}
\end{equation}

In this equation, the total Lagrangian is $L_{tot}=L_{OC}+L_{MR}+L_{MC}+L_{OC-MR}$ where $L_{OC}$, $L_{MR}$, $L_{MC}$, and $L_{OC-MR}$ are the optical cavity, microresonator, microcavity, and the optical cavity - the microrsonator interaction Lagrangian, respectively. In addition, the microwave cavity and the Lagrangian microresonator interaction are defined by $C(x)$ in $L_{MC}$. In Eq. \ref{eq:cavity_Lagrangian}, $\alpha_C$, $\omega_m$, $\omega_C$, $L$, $V_d$, $C(x)$, $C_d$ and $m$ are the coupling coefficient between the OC and the MR, the angular frequency of the oscillation of the MR, the resonance frequency of the optical cavity, the inductance of the microwave circuit, the external driving field applied to the MC and the components of the LC circuit, namely the variable and fixed capacitors, as well as the effective mass of the mechanical resonator. Furthermore, the symbols $A$, $X$, and $\phi$ denote the vector potential, the mechanical resonator position operator, and the magnetic flux operator, respectively, each of which plays a critical role in the quantum dynamics of the system. In the subsequent step, the Hamiltonian formalism must be developed to fully describe the energy dynamics of the system. This involves identifying the conjugate momentum operators corresponding to the canonical variables $A$, $X$, and $\phi$, based on the classical definition of conjugate variables \cite{aggarwal2014selective}. Once the canonical pairs are established, the creation ($\hat{a}^\dagger$) and annihilation ($\hat{a}$) operators are introduced to quantize the fields. These operators enable a compact and tractable representation of the system’s quantum Hamiltonian, which is then rewritten in the operator formalism as:

\begin{equation}
    \begin{array}{l}
        H_{O C}=\hbar \omega_{c} \hat{a}_{c}^{\dagger} \hat{a}_{c} \\[6pt]
        H_{M R}=\frac{\hbar \omega_{m}}{2}\left(\hat{p}_{x}^{2}+\hat{q}_{x}^{2}\right) \\[6pt]
        H_{M C}=\hbar \omega_{\omega} \hat{c}_{\omega}^{\dagger} \hat{c}_{\omega} - j V_{d} C_{d} \sqrt{\frac{\hbar \omega_{\omega}}{2 C_{t}}}\left(\hat{c}_{\omega} - \hat{c}_{\omega}^{\dagger}\right) \\[6pt]
        H_{O C-M R}=\hbar \sqrt{\frac{\alpha_{c}^{2} \omega_{m}}{2 \varepsilon_{0} \omega_{c} m}}\left(\hat{a}_{c}^{\dagger}+\hat{a}_{c}\right) \hat{p}_{x} \\[6pt]
        H_{M C-M R}=C_{p} C_{t} \frac{\hbar \omega_{\omega}}{2} \sqrt{\frac{\hbar}{\omega_{m} m}} \hat{q}_{x}\left(\hat{c}_{s}-\hat{c}_{s}^{\dagger}\right)^{2} \\[6pt]
        H_{O C-\text { drive }}=i \hbar E_{c}\left(\hat{a}_{c}^{\dagger} e^{\left(-j \omega_{L} t\right)}-\hat{a}_{c} e^{\left(j \omega_{L} t\right)}\right)
    \end{array}
    \label{eq:cavity_Hamiltonian}
\end{equation}

In Eq. \ref{eq:cavity_Hamiltonian}, ($\hat{a}_{c}^{\dagger}$, $\hat{a}_{c}$) and ($\hat{c}_{\omega}^{\dagger}$, $\hat{c}_{\omega}$) the optical and microwave cavities contribute to the creation and annihilation operators, respectively. In addition, $E_c$ and $\hbar$ are the input driving rate of the optical cavity and the reduced Planck constant, respectively. Moreover, $C_p$ and $C_t$ are derived using $C_{p}=C^{\prime}(x) /\left[C(x)+C_{d}\right]^{2}$ and $C_t=C_d+C(x_0)$, where $x_0$ denotes the equilibrium position of the MR. In Eq. \ref{eq:cavity_Hamiltonian}, the second term within the microwave cavity Hamiltonian $H_{MC}$ represents the influence of an externally applied driving field on the cavity. Importantly, this expression explicitly includes the interaction Hamiltonians that describe the coupling between the OC and the MR, as well as between the MC and the MR. These terms are critical because they reveal how the coupling between subsystems is mediated via the mechanical resonator. By adjusting the relevant parameters, such as coupling strengths, detuning frequencies, or drive powers, one can dynamically control the interaction strengths between the cavities. This, in turn, provides a direct mechanism for modulating the degree of entanglement between the optical and microwave modes, a key requirement for efficient quantum transduction and hybrid quantum network applications. The equations of motion for this system can be systematically derived using the Heisenberg-Langevin formalism \cite{salmanogli2020entanglement, salmanogli2021design}, which incorporates both unitary evolution and open system dynamics. An essential aspect of this approach is accounting for the interaction between the system and its environment, which introduces dissipation and quantum noise. These are modeled using input noise operators and damping terms, allowing for a realistic description of decoherence. Once the environmental effects are properly included, the complete set of quantum Langevin equations can be written as:

\begin{equation}
    \begin{array}{l}
        \dot{a_{s}}=-\left(j \Delta_{c}+\kappa_{c}\right) \hat{a}_{c}-j G_{1} \hat{p}_{x}+E_{c}+\sqrt{2 \kappa_{s}} \hat{a}_{i n} \\[6pt]
        \dot{c_{\omega}}=-\left(j \Delta_{\omega}+\kappa_{\omega}\right) \hat{c}_{\omega}+j \Delta_{o \omega 1} G_{2} \hat{q}_{x} \hat{c}_{\omega}+E_{\omega}+\sqrt{2 \kappa_{c s}} \hat{c_{i n}} \\[6pt]
        \dot{q_{x}}=\omega_{m} \hat{p}_{x}+G_{1}\left(\hat{a_{c}}+\hat{a_{c}}\right) \\[6pt]
        \dot{p_{x}}=-\gamma_{m} \hat{p}_{x}-\omega_{m} \hat{q}_{x}+\Delta_{\omega} G_{2} \hat{c}_{\omega}^{\dagger} \hat{c}_{\omega}+\hat{b}_{i n}
    \end{array}
    \label{eq:cavity_langevin}
\end{equation}

In this equation, $\kappa_c$, $\gamma_m$, $\kappa_w$, $\Delta_c$, and $\Delta_{\omega}$ are the damping rate of OC, MR and MC and the related de-tuning frequencies, respectively. Also, $G_1 = \sqrt{\alpha_c^2 \omega_m / 2 \varepsilon_0 m \omega_c}$, $G_2 = C_p C_t \sqrt{\hbar / 
_{s}}$, and $E_{\omega} = V_d C_d \sqrt{w_w/2 \hbar C_t}$. The optical cavity driving field is given as $E_c = \sqrt{2 P_c \kappa_c \hbar \omega_L}$, where $P_c$ represents the excitation power applied to drive the cavities \cite{salmanogli2019entanglement, barzanjeh2011entangling}. Eq. \ref{eq:cavity_langevin} is a non-linear operator equation, which generally cannot be solved analytically in terms of the full quantum operators of the system due to its complex interaction terWs and quantum noise contributions. In order to proceed, a common and effective technique is to linearize the dynamics around a steady-state operating point. This is achieved by assuming the system is driven by a strong coherent field, which allows one to decompose each operator into its classical steady-state mean (DC component) and small quantum fluctuations around that mean. Such an approximation is valid when the driving field is sufficiently strong, causing the quantum field dynamics to remain close to the classical operating point. This method, often referred to as linearization around the fixed point, is widely used in cavity optomechanics and quantum transduction analyzes \cite{salmanogli2019entanglement, vitali2007optomechanical}. Consequently, each cavity mode operator (for both optical and microwave fields) can be rewritten as $\boldsymbol{a}_c = A_s + \delta \boldsymbol{a}_c$, $\boldsymbol{c}_w = C_s + \delta \boldsymbol{c}_{\omega}$, $\boldsymbol{q}_x = X_s + \delta \boldsymbol{q}_x$, and $\boldsymbol{P}_x = p_x + \delta \boldsymbol{p}_x$, where capitalized variables with the subscript $"s"$ denote the steady-state (DC) solutions of the system, i.e., the operating point values when the cavities are driven by a continuous external field. Delta-prefix operators, such as $\delta a(t)$, represent the quantum fluctuations around these fixed-point solutions. By substituting these linearized expressions into the nonlinear equation (\ref{eq:cavity_langevin}), and retaining only the first-order terms in the fluctuation operators (i.e., neglecting higher-order nonlinear fluctuation terms), one obtains a linearized set of equations that describe the dynamics of the system near the steady state. This approach isolates the stationary regime behavior, allowing analytical treatment of quantum noise, stability, entanglement generation, and signal transduction characteristics. The resulting linearized equations—commonly referred to as the quantum Langevin equations in fluctuation form — are foundational for studying quantum correlations and noise spectra in driven - dissipative hybrid systems. These expressions form the basis for calculating key performance metrics such as squeezing levels, entanglement rates, and transduction efficiency. Then Eq. \ref{eq:cavity_langevin} is expressed as:

\begin{equation}
    \begin{array}{l}
        -\left(j \Delta_{c}+\kappa_{c}\right) A_{s}-j G_{1} P_{s}+E_{c}=0 \\[6pt]
        -\left(j \Delta_{\omega}+\kappa_{\omega}\right) C_{s}+j \Delta_{o \omega 1} G_{2} X_{s} C_{s}+E_{\omega}=0 \\[6pt]
        \omega_{m} P_{s}+2 G_{1} \operatorname{Re}\left\{A_{s}\right\}=0 \\[6pt]
        -\gamma_{m} P_{s}-\omega_{m} X_{s}+\Delta_{\omega} G_{2}\left|C_{s}\right|^{2}=0
    \end{array}
    \label{eq:cavity_langevin2}
\end{equation}

\vspace{2em}

In order to solve Eq.~\ref{eq:cavity_langevin2}, it is typically assumed that $\{A_s\} \gg 1$  and $|C_s| \gg 1$.  The steady-state values of the system variables - $A_s$, $C_s$, $P_s$, and $X_s$ - are then determined accordingly:

\begin{equation}
    \begin{array}{l}
        P_{s}=\frac{2 G_{1} \operatorname{Re}\left\{A_{s}\right\}}{\omega_{m}}, X_{s}=\frac{-\gamma_{m} P_{s}+\Delta_{\omega} G_{2}\left|C_{s}\right|^{2}}{\omega_{m}} \\ [1em]
        A_{s}=\frac{E_{c}-j G_{1} P_{s}}{\left(j \Delta_{c}+\kappa_{c}\right)}, C_{s}=\frac{j \Delta_{o \omega 1} G_{2} X_{s} C_{s}+E_{\omega}}{\left(j \Delta_{\omega}+\kappa_{\omega}\right)}
    \end{array}
    \label{eq:cavity_langevin3}
\end{equation}

Equation \ref{eq:cavity_langevin3} specifies the steady-state conditions, also known as DC operating points, at which the cavities are subject to external driving. Our analysis that follows is primarily concerned with examining small perturbations occurring around these equilibrium states. It can be demonstrated without difficulty that the selection of DC operating points in Equation \ref{eq:cavity_langevin3} does not affect the behavior of these perturbations. Therefore, the differential equations describing the dynamics of the mode perturbations in proximity to these steady-state values, represented by $(A_s, P_s, X_s, \text{and } C_s)$, are formulated as follows:
 
\begin{widetext}
    \begin{equation}
    \begin{array}{l}
        \dot{\delta a_{s}}=-\left(j \Delta_{c}+\kappa_{c}\right) \hat{\delta} a_{c}-i G_{1} \hat{\delta} p_{x}+\sqrt{2 \kappa_{s}} \hat{\delta} a_{i n} \\[6pt]
        \dot{\delta c_{\omega}}=i \Delta_{\omega 1} G_{2}\left\{\hat{\delta} q_{x} C_{s}+X_{s} \hat{\delta} c_{\omega}\right\}+\sqrt{2 \kappa_{c s}} \hat{\delta} c_{i n}-\left(i \Delta_{\omega}+\kappa_{\omega}\right) \hat{\delta} c_{\omega} \\[6pt]
        \dot{\delta q_{x}}=\omega_{m} \hat{\delta} p_{x}+G_{1}\left(\hat{\delta} a_{c}+\hat{\delta} a_{c}\right) \\[6pt]
        \dot{\delta p_{x}}=-\gamma_{m} \hat{\delta} p_{x}+\Delta_{\omega} G_{2}\left\{\hat{\delta} c_{\omega}{ }^{+} C_{s}+C_{s}^{*} \hat{\delta} c_{\omega}\right\}-\omega_{m} \hat{\delta} q_{x}+\hat{\delta} b_{i n}
\end{array}
\label{eq:cavity_langevin4}
\end{equation}
\end{widetext}

In Eq. \ref{eq:cavity_langevin4} represents the linearized set of coupled equations that describe the quantum fluctuations of the cavity modes. By solving these equations, one can analyze both the separability and the degree of entanglement among the modes. The interaction between the cavity modes in the converter facilitates the generation of CV entanglement, which manifests itself as quantum correlations between the quadrature operators of the intra-cavity fields \cite{shih2003entangled, salmanogli2018raman, salmanogli2019modification}. This entanglement is fundamentally governed by the coupling coefficients, which can be engineered to tailor the correlation strength and structure among the cavity modes. In particular, Eq. \ref{eq:cavity_langevin4} shows that the fluctuation $\delta p_x$ couples to $\delta a_c$ through the coupling constant $G_1$, $\delta w_c$ is influenced by $\delta q_x$ through the term $C_s G_2 \Delta w_1$, $\delta q_x$ driven by changes in $\delta X_c$, and finally $\delta p_x$ is modulated by both $\delta q_x$ and $\delta w_c$. It is therefore the responsibility of the quantum radar system designer to appropriately select and tune these coupling parameters to establish the desired quantum correlations between modes. The influence of these parameters on the degree of mode entanglement will be thoroughly investigated in the following section. Note that in Eq. \ref{eq:cavity_langevin4}, the OC and MC modes are represented solely using annihilation operators, whereas for the MR, the quadrature operators are used. This distinction enables the straightforward derivation of the quadrature representations for the OC and MC modes. Accordingly, the full system dynamics can be compactly expressed using the QED formalism in matrix form:

\begin{widetext}
\begin{equation}
\left[
\begin{array}{c}
\dot{\delta q_x} \\
\dot{\delta p_x} \\
\dot{\delta X_c} \\
\dot{\delta Y_c} \\
\dot{\delta X_\omega} \\
\dot{\delta Y_\omega}
\end{array}
\right]
=
\underbrace{
\left[
\begin{array}{cccccc}
0 & \omega_m & G_1 \sqrt{2} & 0 & 0 & 0 \\
-\omega_m & -\gamma_m & 0 & G_m & 0 & 0 \\
0 & 0 & -\kappa_c & \Delta_c & 0 & 0 \\
0 & -G_1 \sqrt{2} & -\Delta_c & -\kappa_c & 0 & 0 \\
G_{11} & 0 & 0 & 0 & -\kappa_{\omega 1} & \Delta_{\omega 1} \\
G_{22} & 0 & 0 & 0 & -\Delta_{\omega 1} & -\kappa_{\omega 1}
\end{array}
\right]
}_{A_{i,j}}
\times
\underbrace{
\left[
\begin{array}{c}
\delta q_x \\
\delta p_x \\
\delta X_c \\
\delta Y_c \\
\delta X_\omega \\
\delta Y_\omega
\end{array}
\right]
}_{\mathbf{u}(0)}
+
\underbrace{
\left[
\begin{array}{c}
0 \\
\delta b_{in} \\
\sqrt{2\kappa_c} \, \delta X_c^{\text{in}} \\
\sqrt{2\kappa_c} \, \delta Y_c^{\text{in}} \\
\sqrt{2\kappa_\omega} \, \delta X_\omega^{\text{in}} \\
\sqrt{2\kappa_\omega} \, \delta Y_\omega^{\text{in}}
\end{array}
\right]
}_{\mathbf{n}(t)}
\label{eq:cavity_langevin5}
\end{equation}
\end{widetext}

The parameters used in this equation are defined as $G_m = \sqrt{2G_2 \Delta_{\omega}C_s}$, $G_{11} = -\sqrt{2G_2\Delta_{\omega}\operatorname{Im}\{C_s\}}$, $G_{22} = \sqrt{2G_2\Delta_{\omega}\operatorname{Re}\{C_s\}}$, $\kappa_{\omega 1}=\kappa_{\omega}+G_{2} \Delta_{\omega} \operatorname{Im}\left\{q_{s}\right\}$, and $\Delta_{\omega 1}=\Delta_{\omega}-\mathrm{G}_{2} \Delta_{\omega} \operatorname{Re}\left\{\mathrm{q}_{\mathrm{s}}\right\}$. One can consider the solution of Eq. \ref{eq:cavity_langevin5}, which is $u(t) = e^{(A_{i,j}t)}u(0) + \int e^{(A_{i,j}s)}n(t-s)ds$, where $n(.)$ defines the noise of the system. The familiar correlation function can be applied to characterize the noises in the system as \cite{shih2003entangled, salmanogli2019entanglement, salmanogli2020optical, salmanogli2021design}:  

\begin{widetext}
\begin{equation}
    \begin{array}{l}
    <a_{i n}(s) a_{i n}^{*}\left(s^{\prime}\right)>=\left[N\left(\omega_{c}\right)+1\right] \delta\left(s-s^{\prime}\right) ;<a_{i n}^{*}(s) a_{i n}\left(s^{\prime}\right)>=\left[N\left(\omega_{c}\right)\right] \delta\left(s-s^{\prime}\right) \\ [6pt]
    <c_{i n}(s) c_{i n}^{*}\left(s^{\prime}\right)>=\left[N\left(\omega_{\omega}\right)+1\right] \delta\left(s-s^{\prime}\right) ;<c_{i n}^{*}(s) c_{i n}\left(s^{\prime}\right)>=\left[N\left(\omega_{\omega}\right)\right] \delta\left(s-s^{\prime}\right) \\[6pt]
    <b_{i n}(s) b_{i n}^{*}\left(s^{\prime}\right)>=\left[N\left(\omega_{m}\right)+1\right] \delta\left(s-s^{\prime}\right) ;<b_{i n}^{*}(s) b_{i n}\left(s^{\prime}\right)>=\left[N\left(\omega_{m}\right)\right] \delta\left(s-s^{\prime}\right)
    \end{array}
\label{eq:cavity_langevin6}
\end{equation}
\end{widetext}

In Eq. \ref{eq:cavity_langevin6}, $\left[e^{(\hbar w/k_B T)}-1\right]^{-1}$, where $T$ and $k_B$ denote the operational temperature and the Boltzmann constant, respectively. This function $N(\omega)$ represents the mean number of thermal photons in thermal equilibrium for a mode of frequency $\omega$. In Eq. \ref{eq:cavity_langevin5} describes the fluctuations of the cavity modes and forms the basis for analyzing mode entanglement. In the context of quantum radar applications, special emphasis is placed on the entanglement between the OC and MC modes, as it plays a critical role in the overall performance and sensitivity of the system. Accordingly, our focus is directed toward quantifying and characterizing the correlation between the OC and MC modes. To achieve this, all potential factors that may degrade or destroy the entanglement between modes are thoroughly examined. The degree of entanglement is assessed using the Simon-Peres-Horodecki criterion, which provides a necessary and sufficient condition for CV bipartite entanglement \cite{horodecki1996separability,peres1996separability, simon2000peres, tasgin2020measuring}: 

\begin{widetext}
\begin{equation}
    \lambda_{S P H}=\operatorname{det}(A) \cdot \operatorname{det}(B)+(0.25-|\operatorname{det}(C)|)^{2}-\operatorname{tr}\left(A J C J B J C^{T} J\right)-0.25 \times(\operatorname{det}(A)+\operatorname{det}(B)) \geq 0
    \label{eq:cavity_langevin7}
\end{equation}
\end{widetext}

In Eq. \ref{eq:cavity_langevin7}, $J$ is a matrix like $J = \left[0, 1; -1, 0\right]$, and "$tr$" represents the matrix trace operation. The criterion presented in Eq. \ref{eq:cavity_langevin7} serves as a necessary and sufficient condition for the separability of bipartite Gaussian states. Consequently, the analysis assumes that all the cavity states considered exhibit Gaussian statistics. In Eq. \ref{eq:cavity_langevin8}, the matrices A, B and C represent the local covariance matrices and the cross-correlation matrix, respectively, forming the full covariance matrix in the block structure $J = \left[A, C; C^T, B\right]$. The elements of the mentioned matrix are exemplified for the OC-MC modes as:

\begin{widetext}
\begin{equation}
\begin{array}{l}
    A=\resizebox{.9\textwidth}{!}{$\left[\begin{array}{cc}
    <\delta X_{c}^{2}>-<\delta X_{c}>^{2} & 0.5 \times<\delta X_{c} \delta Y_{c}+\delta Y_{c} \delta X_{c}>-<\delta X_{c}><\delta Y_{c}> \\
    0.5 \times<\delta X_{c} \delta Y_{c}+\delta Y_{c} \delta X_{c}>-<\delta X_{c}>\delta Y_{c}> & <\delta Y_{c}^{2}>-<\delta Y_{c}>^{2}
    \end{array}\right]$} \\[1em]
    
    B=\resizebox{.9\textwidth}{!}{$\left[\begin{array}{cc}
    <\delta X_{\omega}^{2}>-<\delta X_{\omega}>^{2} & 0.5 \times<\delta X_{\omega} \delta Y_{\omega}+\delta Y_{\omega} \delta X_{\omega}>-<\delta X_{\omega}><\delta Y_{\omega}> \\
    0.5 \times<\delta Y_{\omega} \delta X_{\omega}+\delta X_{\omega} \delta Y_{\omega}>-<\delta Y_{\omega}><\delta X_{\omega}> & <\delta Y_{\omega}^{2}>-<\delta Y_{\omega}>^{2}
    \end{array}\right]$} \\[1em]

    C=\resizebox{.9\textwidth}{!}{$\left[\begin{array}{cc}
    0.5 \times<\delta X_{c} \delta X_{\omega}+\delta X_{\omega} \delta X_{c}>-<\delta X_{c}><\delta X_{\omega}> & 0.5 \times<\delta X_{c} \delta Y_{\omega}+\delta Y_{\omega} \delta X_{c}>-<\delta X_{c}>\delta Y_{\omega}> \\
    0.5 \times<\delta Y_{c} \delta X_{\omega}+\delta X_{\omega} \delta Y_{c}>-<\delta Y_{c}>\delta X_{\omega}> & 0.5 \times<\delta Y_{c} \delta Y_{\omega}+\delta Y_{\omega} \delta Y_{c}>-<\delta Y_{c}><\delta Y_{\omega}>
    \end{array}\right]$}
\end{array}
\label{eq:cavity_langevin8}
\end{equation}
\end{widetext}

For each two-mode entanglement evaluation, we need to construct matrix elements and use criterion expressed in Eq. \ref{eq:cavity_langevin7} to analyze the correlation between modes. For doing so, the calculation of the quadrature modes operators expectation value ($\left\langle\delta \mathrm{X}_{\mathrm{c}}\right\rangle,\left\langle\delta \mathrm{X}_{\omega}\right\rangle,\left\langle\delta \mathrm{X}_{\mathrm{c}}^{2}\right\rangle,\left\langle\delta \mathrm{X}_{\omega}^{2}\right\rangle$) are necessary to study the modes entanglement. 

\begin{figure*}[htbp]
    \centering
    % Row 1: (a) and (b)
    \begin{subfigure}[b]{0.48\textwidth}
        \centering
        \includegraphics[width=\linewidth]{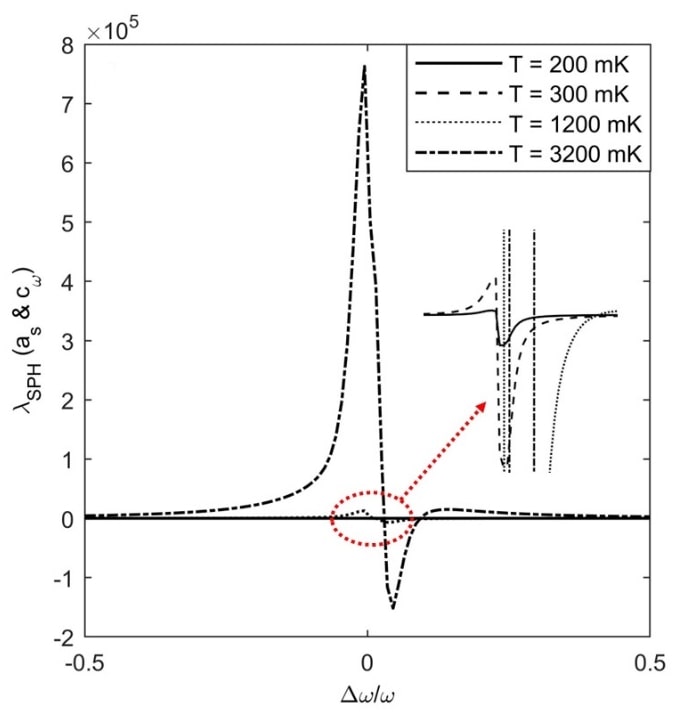}
        \caption{}
        \label{fig:q_entg_betw_cavity_modes_a}
    \end{subfigure}
    \hfill
    \begin{subfigure}[b]{0.5\textwidth}
        \centering
        \includegraphics[width=\linewidth]{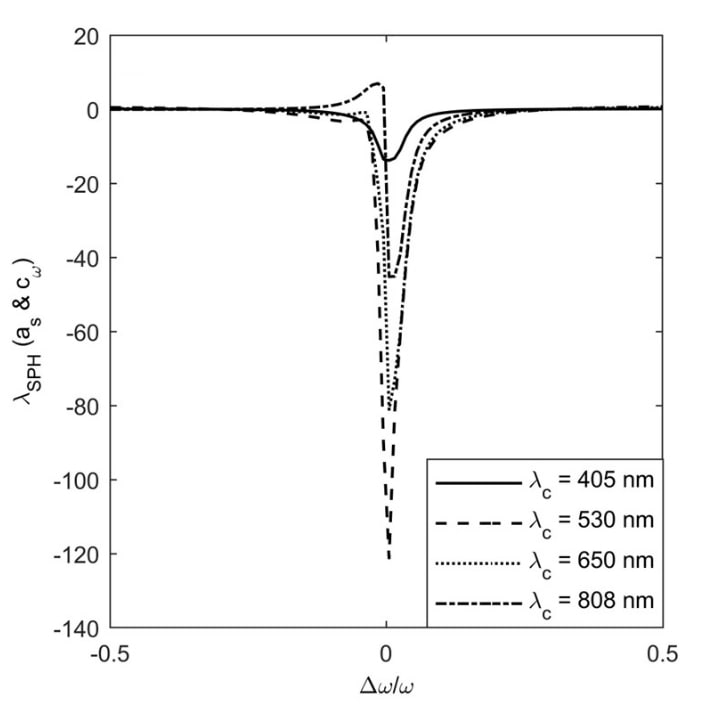}
        \caption{}
        \label{fig:q_entg_betw_cavity_modes_b}
    \end{subfigure}

    \vspace{0.5cm}

    % Row 2: (c) and (d)
    \begin{subfigure}[b]{0.48\textwidth}
        \centering
        \includegraphics[width=\linewidth]{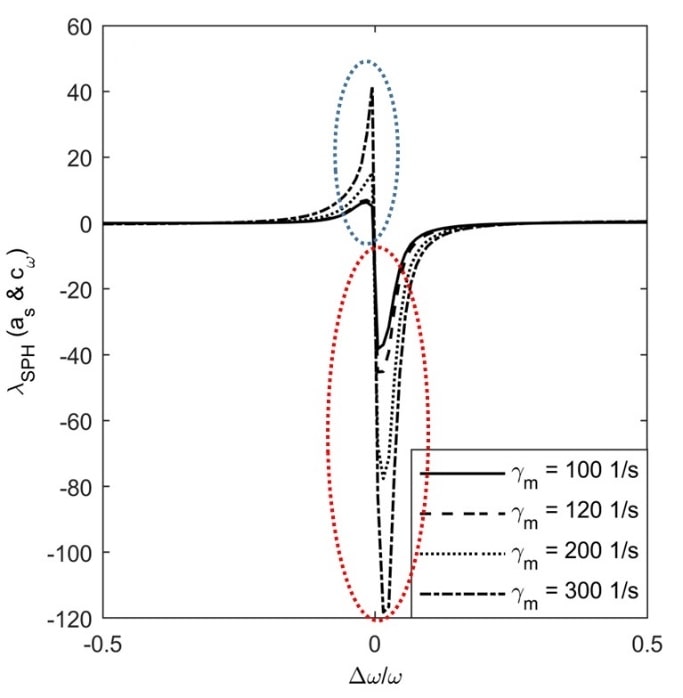}
        \caption{}
        \label{fig:q_entg_betw_cavity_modes_c}
    \end{subfigure}
    \hfill
    \begin{subfigure}[b]{0.5\textwidth}
        \centering
        \includegraphics[width=\linewidth]{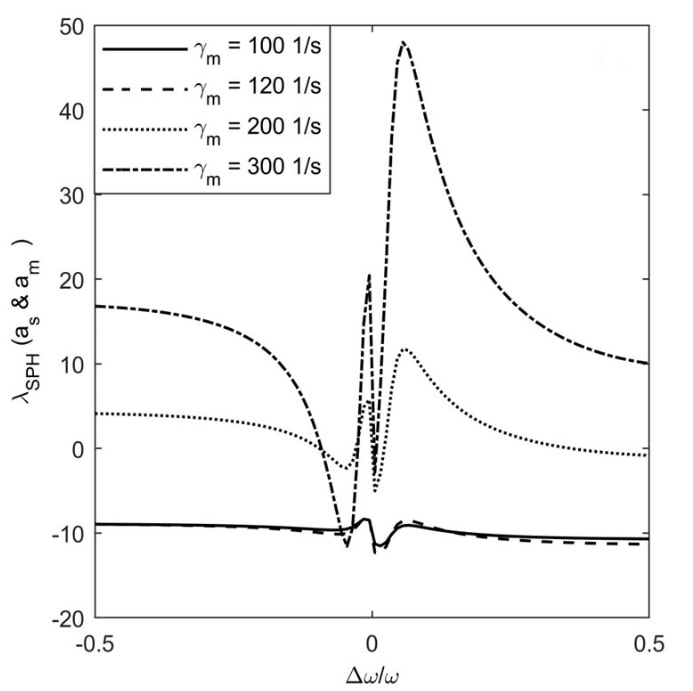}
        \caption{}
        \label{fig:q_entg_betw_cavity_modes_d}
    \end{subfigure}

    \caption{Entanglement between cavities modes; $C_s$ and $q_s$ are real under varying parameters: (a) Temperature effect, (b) incident source wavelength effect, (c), and (d) study of the effect of the MR \cite{salmanogli2020entanglement}.}
    \label{fig:composite}
\end{figure*}

The influence of key parameters in the tripartite system, namely MR damping rate, incident wavelength, and temperature, on the entanglement between cavity modes is systematically investigated. The impact of temperature on the system is presented in Fig. \ref{fig:q_entg_betw_cavity_modes_a}. As expected, increasing the system's temperature leads to a significant reduction in entanglement, ultimately causing the cavity modes to become separable. Detailed behavior at 200 mK and 300 mK is highlighted in the magnified inset.

A critical factor limiting the operation of the tripartite system to cryogenic temperatures is the relatively low resonance frequency of the MR cavity ($w_m = 2 \pi \times 10^6 $). According to Eq. \ref{eq:cavity_langevin6} thermal noise becomes dominant at such low frequencies, severely degrading entanglement. To address this issue, several approaches have been proposed, including frequency band engineering and replacing the MR with a high-frequency optoelectronic component \cite{salmanogli2020entanglement}. Therefore, to maintain entanglement, the operational temperature should remain below approximately 1200 mK. Another crucial parameter affecting entanglement is the wavelength of the excitation source. As shown in Fig. \ref{fig:q_entg_betw_cavity_modes_b}, varying the incident wavelength modulates the degree of entanglement between the cavity modes. In Fig. \ref{fig:q_entg_betw_cavity_modes_c}, the effect of the MR damping rate on the cavity modes of the OC and MC is analyzed, revealing that the entanglement is highly sensitive to this rate. This sensitivity arises from the role of the MR oscillator in mediating the coupling between the optical and microwave cavities. For further insight, Fig. \ref{fig:q_entg_betw_cavity_modes_d} shows the entanglement between the MR and OC modes as a function of the MR damping rate. As the damping increases, the coupling between MR and OC weakens, leading to a pronounced deterioration in the entanglement.

Thus far, the dynamics of the EOM converter have been analytically derived using the canonical quantization method in conjunction with the Heisenberg-Langevin formalism. However, in a practical quantum radar scenario, the system must be able to transmit entangled microwave photons toward a remote target. This implies that the entangled photons will propagate through the atmosphere before interacting with the target. Naturally, similar to classical radar systems, signal dispersion and degradation occur during atmospheric propagation. To counteract this loss, an amplifier may be used to boost the signal strength prior to atmospheric transmission. However, amplifying entangled photons while preserving their quantum correlations presents a significant challenge. Quantum amplification must avoid introducing excessive noise that could degrade the non-classical correlations. Furthermore, after interacting with the target, the scattered photons once again traverse the atmospheric channel, may be amplified, and are ultimately detected. In the next stage of this work, another quantum device that can be used to generate non-classicality will be discussed in detail using quantum theory, along with the other criterion to analyze the quantum entanglement generated. 

\subsubsection{Opto-Electronic Converter System Definition: Symplectic Eigenvalue Criterion (\texorpdfstring{$2\eta$}{2eta})}

\begin{figure*}[htbp]
    \centering
    % Row 1: (a) and (b)
    \begin{subfigure}[b]{0.40\textwidth}
        \centering
        \includegraphics[width=\linewidth]{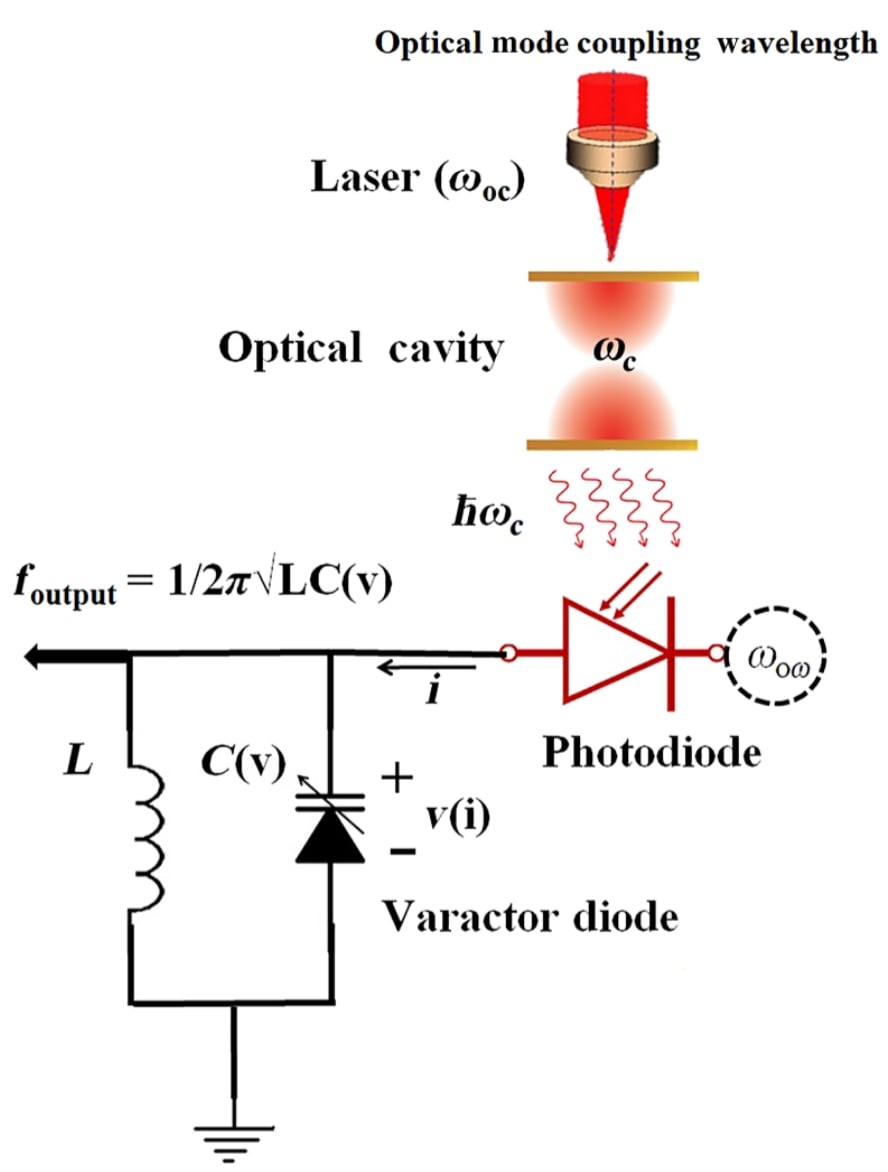}
        \caption{}
        \label{fig:q_optoelectronic_system_schematic_a}
    \end{subfigure}
    \hfill
    \begin{subfigure}[b]{0.45\textwidth}
        \centering
        \includegraphics[width=\linewidth]{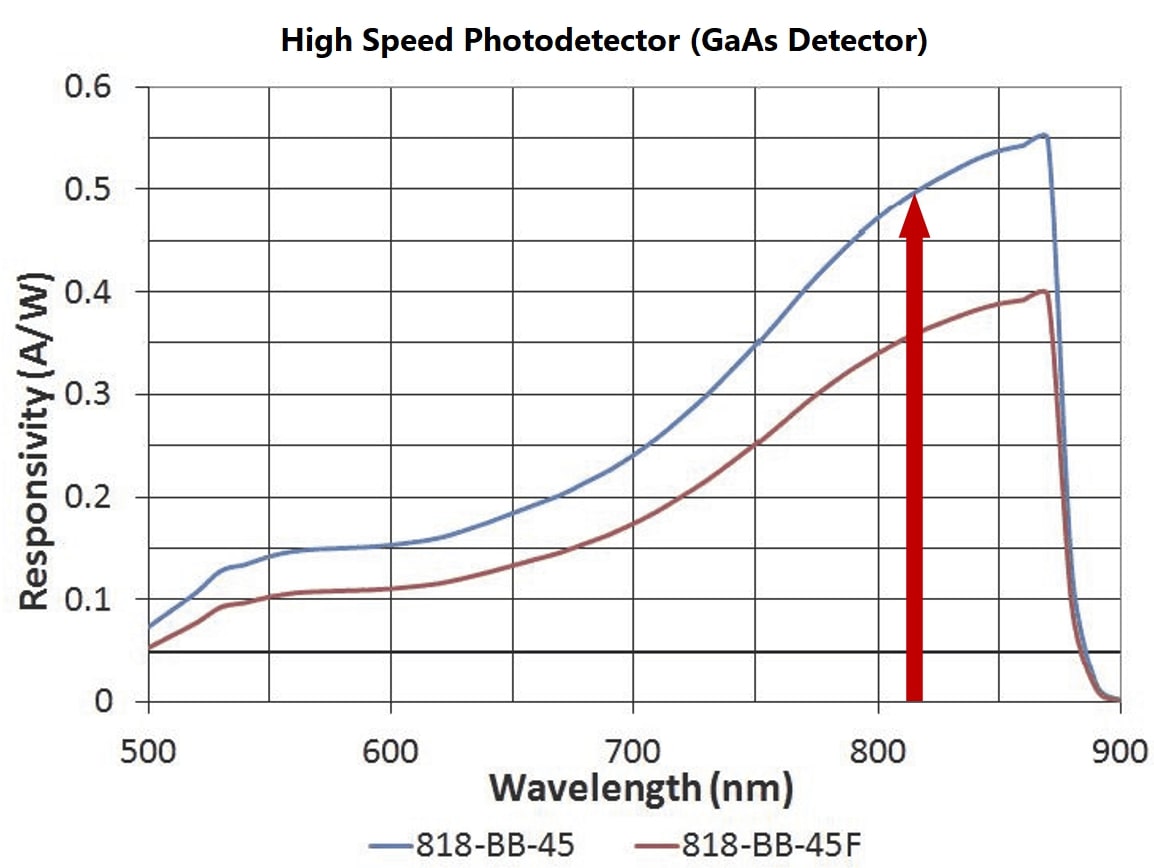}
        \caption{}
        \label{fig:q_optoelectronic_system_schematic_b}
    \end{subfigure}

    \vspace{0.5cm}

    % Row 2: (c) and (d)
    \begin{subfigure}[b]{0.36\textwidth}
        \centering
        \includegraphics[width=\linewidth]{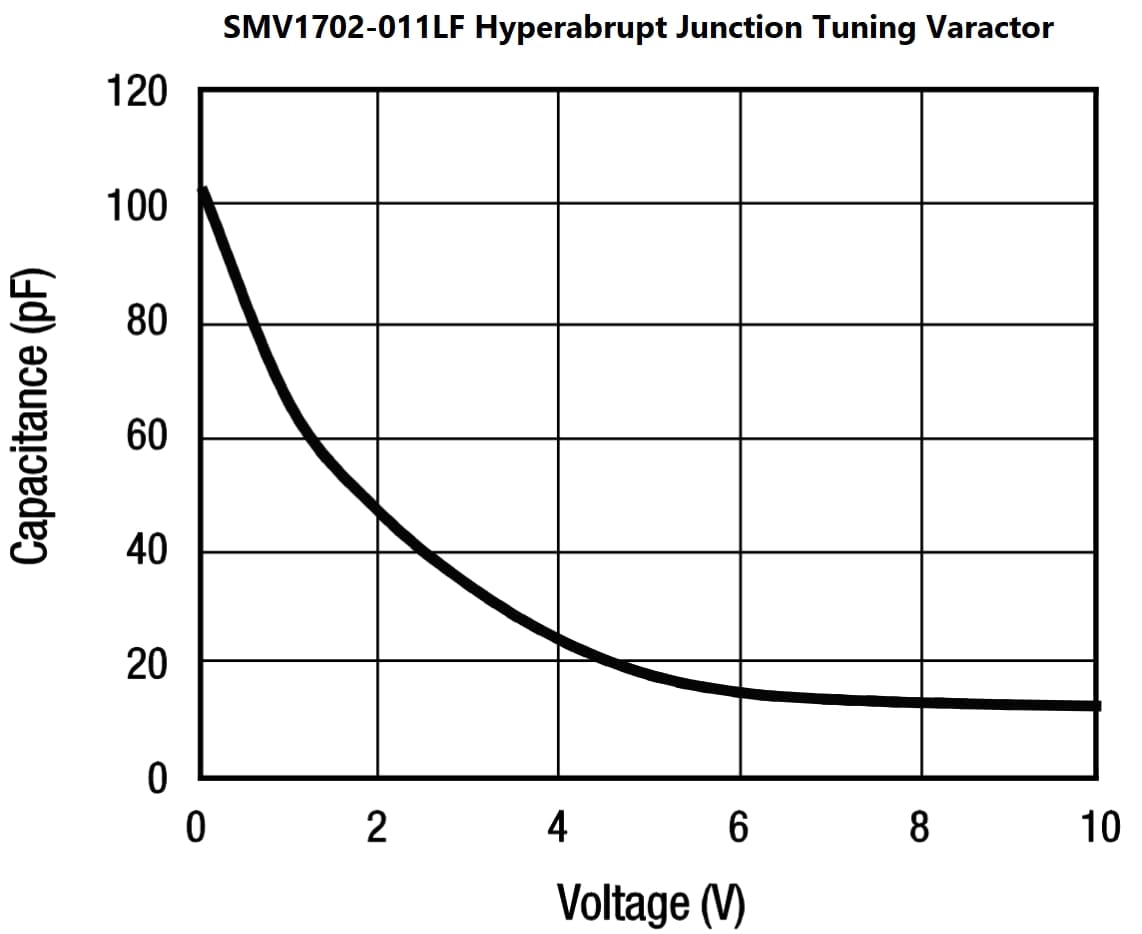}
        \caption{}
        \label{fig:q_optoelectronic_system_schematic_c}
    \end{subfigure}
    \hfill
    \begin{subfigure}[b]{0.65\textwidth}
        \centering
        \includegraphics[width=\linewidth]{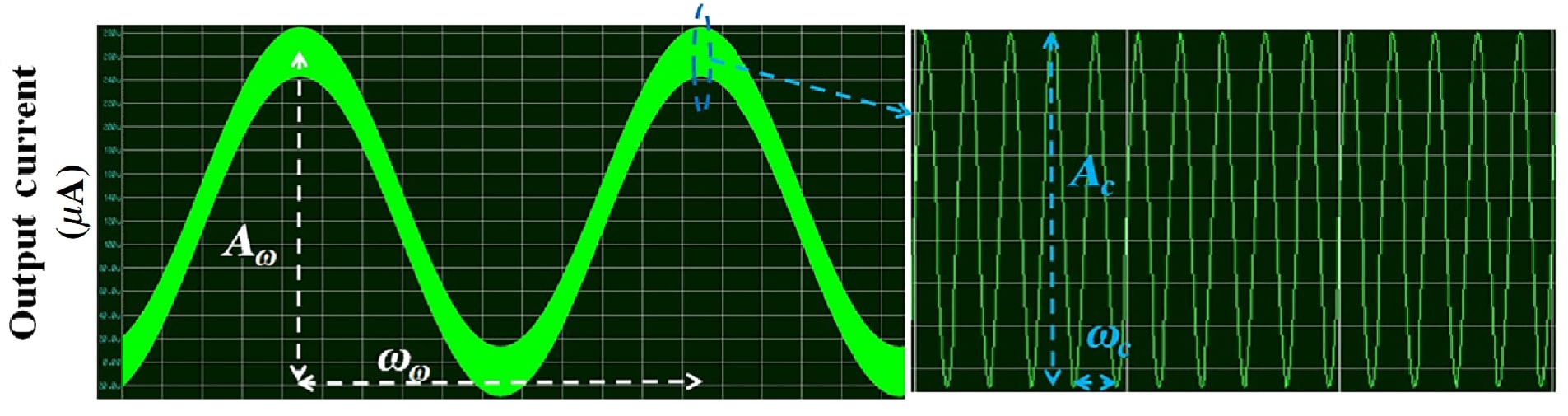}
        \caption{}
        \label{fig:q_optoelectronic_system_schematic_d}
    \end{subfigure}

    \caption{Optoelectronic system schematic; $C_s$ and $q_s$ are real under varying parameters: (a) OC mode coupling to MC mode through the photodetector and a Varactor diode, (b) a typical GaAs photodetector responsivity graph \cite{newport_biased_detector}, (c) Varactor diode capacitance variation vs biased voltage, which is the function of photodetector current \cite{junction_tuning_varactor} and (d) simulated photocurrent as a function of optical cavity mode incidence wave frequency and amplitude. The optical mode center wavelength is around 808 nm \cite{salmanogli2019entanglement}.}
    \label{fig:composite2}
\end{figure*}

The proposed optoelectronic system, depicted in Fig. \ref{fig:composite2}, comprises an optical cavity, a photodetector, a varactor diode (VD), and an inductor \cite{salmanogli2019entanglement}. In this configuration, the OC mode is initially coupled to the photodetector, which generates a photocurrent dependent on the incident photon energy represented as $i = f_1(\hbar \omega)$, where $f_1$ denotes the optical coupling function. The resulting photocurrent is then directed through a varactor diode, leading to a change in its capacitance, described by $C = f_2(\hbar \omega)$, with $f_2$ acting as the electrical coupling factor. This capacitance variation directly influences the resonance frequency of the MC, effectively linking it to the optical input. Thus, the system enables a direct and tunable interaction between the optical and microwave cavity modes via optoelectronic mediation. The photodetector's responsivity characteristics are illustrated in Fig. \ref{fig:q_optoelectronic_system_schematic_b}, while Fig. \ref{fig:q_optoelectronic_system_schematic_c} presents a representative model of the varactor diode, showing how its capacitance varies with the applied voltage. These two figures serve to demonstrate typical device responses and are not tied to a specific implementation. Fig. \ref{fig:q_optoelectronic_system_schematic_d} further clarifies the dynamic interaction chain: the optical cavity couples to the photodetector, which influences the varactor diode through the photocurrent, thus modulating the parameters of the MC.  Here, $A_{\omega}$ represent the amplitude and frequency of the MC mode, respectively, while $\omega _c$ and $A_c$ denote the frequency and amplitude of the OC mode. Crucially, the degree of coupling between the two cavity modes is governed by the tuning of the coupling functions $f_1$ and $f_2$. The ratio $f_1/f_2$ quantifies the extent of the influence of the OC mode on the MC mode. By appropriately engineering this ratio, the system can establish and sustain entanglement between the cavity modes, even under room-temperature conditions, marking a significant advancement in practical quantum information processing and optoelectronic integration. From a system engineering perspective, the converter's ability to sustain and manipulate nonclassical correlations, such as CV entanglement, relies on careful optimization of both coupling paths. In the theoretical analysis that follows, the expressions for these coupling factors are derived and their impact on the entanglement between output modes is examined. In summary, the opto-electronic converter operates as an entanglement interface between the optical and microwave domains. By tuning the opto-electronic and electromagnetic coupling parameters, the degree and nature of the quantum correlations between the OC and MC modes can be precisely engineered to suit the needs of a quantum radar system \cite{salmanogli2019entanglement}. In the following section, a theoretical design of the optoelectronic converter is presented, following a methodology similar to that used in the design of the EOM converter previously discussed. The primary objective is to perform a comprehensive comparison between the two systems with the aim of determining which architecture more effectively maintains entanglement between modes. In addition, strategies are developed to improve non-classical correlations within each system \cite{salmanogli2019entanglement}. Given that a central focus of this research is the realization of sustainable entanglement, it is essential to develop a system architecture that not only facilitates quantum correlations but also preserves them under practical operating conditions. This comparative study will provide critical insight into the advantages and limitations of each type of converter, guiding the optimal design of quantum interfaces for radar and sensing applications. The quantum system incorporating the opto-electronic converter, as illustrated in Fig. \ref{fig:q_optoelectronic_system_schematic_a}, is theoretically analyzed using the canonical quantization method \cite{salmanogli2019entanglement}. Unlike the dipole approximation approach \cite{brandsema2017theoretical, salmanogli2020optoelectronic}, which typically simplifies the interaction between light and matter, the canonical quantization framework emphasizes the full quantum interaction between the incident field and the atomic (or matter-based) system. In general, accurately describing the dynamics of any quantum system requires specifying the total Hamiltonian governing its evolution. Accordingly, the total Hamiltonian for the proposed system is formulated as follows:

\begin{equation}
    \begin{array}{l}
    H_{O C}=\frac{\varepsilon_{0}}{2}\left(E^{2}+\omega_{c}^{2} A^{2}\right) \\ [1em]
    H_{P D}=\frac{P^{2}}{2 m_{e f f}}+\frac{1}{2} m_{e f f} \omega_{e g}^{2} X^{2} \\ [1em]
    H_{M C}=\frac{Q^{2}}{2 C_{0}}+\frac{\phi^{2}}{2 L}+\frac{C_{d} v_{d}}{C_{0}} Q \\ [1em]
    H_{O C-P D}=\alpha_{c} \frac{A P}{m_{e f f}} \\ [1em]
    H_{M C-P D}=\frac{-C^{\prime}(x)}{C_{0}^{2}}\left\{\frac{Q^{2}}{2}+C_{d} v_{d} Q\right\} X
    \end{array}
    \label{eq:q_optical_cavity1}
\end{equation}

In Eq. \ref{eq:q_optical_cavity1}, the operators ($X$, $P$), ($A$, $E$), and ($\phi$, $Q$) represent the position and momentum of the electron-hole pairs in the photodetector (PD), the vector potential and electric field of the optical cavity, and the phase and charge operators associated with the MC, respectively. These constants $\alpha_c$, $\omega_{eg}$, $v_d$, and $C'(x)$, $C_0$, $C_d$, and $m_{eff}$ correspond to the optical-to-microwave cavity coupling coefficient, the PD gap-related transition frequency, the driving voltage of the MC, the position-dependent variable capacitance, the total capacitance, the capacitance between the MC and its driving field, and the effective mass of the electron-hole pair, respectively. To proceed, it is necessary to identify the canonical conjugate variables for $A$, $X$, and $\phi$ to express the Hamiltonian of the system in terms of the creation and annihilation operators. This step follows the classical approach of defining conjugate pairs as outlined in \cite{salmanogli2020optical}. The resulting Hamiltonian, reformulated using these quantized variables, is then given by:

\begin{equation}
    \begin{array}{l}
    H_{O C}=\hbar \omega_{c} \hat{a}_{c}^{\dagger} \hat{a}_{c}+j \hbar E_{c}\left[\hat{a}_{c}^{\dagger} e^{\left(-j \omega_{o c} t\right)}-\hat{a}_{c} e^{\left(j \omega_{o c} t\right)}\right] \\ [1em]
    H_{M R}=\frac{\hbar \omega_{e g}}{2}\left(\hat{p}_{x}^{2}+\hat{q}_{x}^{2}\right) \\ [1em]
    H_{M C}=\hbar \omega_{\omega} \hat{c}_{\omega}^{\dagger} \hat{c}_{\omega}-j\left|v_{d}\right| C_{d} \sqrt{\frac{\hbar \omega_{\omega}}{2 C_{0}}}\left[\hat{c}_{\omega} e^{\left(-j \omega_{o \omega} t\right)}-\hat{c}_{\omega}^{\dagger} e^{\left(j \omega_{o \omega} t\right)}\right] \\ [1em]
    H_{O C-P D}=\hbar \sqrt{\frac{\alpha_{c}^{2} \omega_{e g}}{2 \varepsilon_{0} m_{e f f} \omega_{c}}}\left(\hat{a}_{c}^{\dagger}+\hat{a}_{c}\right) \hat{p}_{x} \\ [1em]
    H_{M C-P D}=-\frac{\hbar \mu_{c} \omega_{\omega}}{2 d} \sqrt{\frac{\hbar}{\omega_{e g} m_{e f f}}} \hat{q}_{x} \hat{c}_{s}^{\dagger} \hat{c}_{s}
    \end{array}
    \label{eq:q_optical_cavity2}
\end{equation}

In Eq. \ref{eq:q_optical_cavity2}, ($a_c^{\dagger}$, $a_c$), and ($c_{\omega}^{\dagger}$, $c_{\omega}$) denote the creation and annihilation operators for the optical and microwave cavity modes, respectively. These parameters $\omega_{oc}$, ($p_x, q_x$), $E_c$, $g_{op}$, and $d$ represent the angular frequency of the input optical source, the normalized quadrature operators, the drive amplitude of the input optical cavity, the optical coupling rate between the cavity and the photodetector, and the width of the VD capacitor depletion layer, respectively. Moreover, $\mu_c = C'(x)/C_0$ captures the sensitivity of the variable capacitance to electron-hole displacement. In the Hamiltonian HMC-PD interaction, which describes the coupling between the microwave cavity and the photodetector, the key coupling strength is given by $\mathrm{g}_{\alpha, \rho}=\left(\mu_{\mathrm{c}} \omega_{\omega} / 2 \mathrm{~d}\right) \times \sqrt{ \left(\hbar / \omega_{\mathrm{eg}} \mathrm{~m}_{\mathrm{eff}}\right)}$. This coupling factor, $g_{ow}$, plays a crucial role in the quantum radar architecture, as it allows for strong manipulation of nonclassical correlations between modes. Notably, by tuning this parameter, it becomes possible to maintain entanglement between the optical and microwave modes even under elevated thermal noise conditions, a point that will be further examined in the Results and Discussion section. The optical coupling rate between the OC and the PD is given similarly by $g_{op} = \sqrt{\left(\omega_{\mathrm{eg}} \alpha_{\mathrm{c}}^{2} / 2 \omega_{\mathrm{c}} \varepsilon_{0} \mathrm{~m}_{\mathrm{eff}}\right)}$. In Eq. \ref{eq:q_optical_cavity2}, the Hamiltonian HMC incorporates both the free evolution of the microwave cavity field and its external drive. One of the key analytical tasks in this framework is the accurate determination of the coupling coefficient between the optical cavity and the photodetector. To this end, within a unit volume, first-order perturbation theory can be used, and the resulting expression is provided as follows \cite{scully1997quantum}:

\begin{widetext}    
\begin{equation}
    g_{o p}=\frac{\pi \omega_{\mathrm{t}}}{\varepsilon_{0} V_{m}} \mu^{2} g_{J}\left(\hbar \omega_{\mathrm{eg}}\right) \cdot L\left(\omega_{\mathrm{eg}}\right) \rightarrow \alpha_{\mathrm{c}}=\sqrt{\frac{2 \omega_{\mathrm{c}} \varepsilon_{0} \mathrm{~m}_{\mathrm{eff}}}{\omega_{\mathrm{eg}}}} \frac{\pi \omega_{\mathrm{c}}}{\varepsilon_{0} V_{m}} \mu^{2} g_{J}\left(\hbar \omega_{\mathrm{eg}}\right) \cdot L\left(\omega_{\mathrm{eg}}\right)
    \label{eq:q_optical_cavity3}
\end{equation}
\end{widetext}

In this expression, $g_J(\hbar \omega_{eg})$, $\mu$, and $L(\omega_{eg})$ represent the density of the states of the photodetector, the dipole moment, and the Lorentzian line-shaped function, respectively. Eq. \ref{eq:q_optical_cavity2} also provides information on the behavior of the coupling constant $\alpha_c$, which is highly sensitive to the electrical and optical properties of the photodetector. Therefore, engineering the converter, particularly the design and material composition of the PD, offers an effective means of modulating the strength of the coupling between the OC and the PD. As an example, plasmonic-based photodetectors can be employed to significantly enhance the optical coupling factor due to their strong light–matter interaction at the nanoscale \cite{scully1997quantum}. Ultimately, this coupling constant serves as a critical design parameter for engineering the interaction between cavity modes and controlling the degree of non-classical correlations, such as entanglement, in the quantum radar system. Following the derivation of the system Hamiltonian, the corresponding equations of motion are formulated using the Heisenberg–Langevin formalism. To accurately capture the real behavior of the system, damping rates and noise terms are incorporated to account for dissipation and fluctuations arising from the interaction between the quantum system and its surrounding environment. Furthermore, applying the rotating wave approximation (RWA), we define the relevant detuning parameters as follows: the detuning between the microwave cavity and its drive is $\Delta_{\omega} = \omega_{\omega} - \omega_{o \omega}$, the optical cavity detuning is $\Delta_c = \omega_c - \omega_{o c}$, and the photodetector detuning is $\Delta_{e g} = \omega_{e g} - \omega_{c}$. With these considerations, the complete set of dynamical equations for the newly proposed opto-electronic converter is established as:

\begin{equation}
    \begin{array}{l}
    \dot{a_{c}}=-\left(j \Delta_{c}+\kappa_{c}\right) \hat{a}_{c}-j g_{o p} \hat{p_{x}} \hat{a}_{c}+E_{c}+\sqrt{2 \kappa_{c}} \hat{a}_{i n} \\ [1em]
    \dot{c_{\omega}}=-\left(j \Delta_{\omega}+\kappa_{\omega}\right) \hat{c_{\omega}}+j g_{\omega p} \hat{a}_{x} \hat{c}_{\omega}+E_{\omega}+\sqrt{2 \kappa_{\omega}} \hat{c}_{i n} \\ [1em]
    \dot{q_{x}}=\Delta_{e g} \hat{p_{x}}+g_{o p}\left(\hat{a}_{c}^{\dagger}+\hat{a}_{c}\right) \\ [1em]
    \dot{p_{x}}=-\gamma_{p} \hat{p}_{x}-\Delta_{e g} \hat{q}_{x}+g_{\omega p} \hat{c}_{\omega}^{\dagger} \hat{c}_{\omega}+\hat{b}_{i n}
    \end{array}
    \label{eq:q_optical_cavity4}
\end{equation}

In Eq. \ref{eq:q_optical_cavity4}, the parameters $\kappa_c$, $\gamma_p$, and $\kappa_{\omega}$ are the damping rates of the optical cavity, photodetector, and microwave cavity, respectively. Furthermore, $\Delta_c$, $\Delta_{\omega}$, and $\Delta_{eg}$ are the related detuning frequencies, and $v_d C_d \sqrt{\left(\omega_{\omega} / 2 \hbar \mathrm{C}_{0}\right)}$ and $\mathrm{E}_{\mathrm{c}}=\sqrt{\left(2 \mathrm{P}_{\mathrm{c}} \kappa_{\mathrm{c}} / \hbar \omega_{o c}\right)}$, where $P_c$ is the OC excitation power \cite{salmanogli2019entanglement, barzanjeh2011entangling, aggarwal2014selective, vitali2007optomechanical}. In addition, $b_{in}$ is the quantum noise attributed to PD acting on it and ain and $c_{in}$ are input noises due to the fact that cavities interact with the environment. In the following analysis, we focus on evaluating the entanglement between the OC and MC as continuous-mode quantum fluctuations around their fixed operating points, where both cavities are externally driven. Under these steady state conditions, the system's equations of motion can be linearized by expanding around the classical (steady state) fields, which serve as fixed points \cite{salmanogli2019entanglement, vitali2007optomechanical}. To carry out this linearization, each cavity mode is expressed as a sum of a steady-state component and a small fluctuation term, as follows: $a_c = A_s + \delta_{ac}$, $c_{\omega} = C_s + \delta{c \omega}$, $q_x = X_s+\delta q_x$, and $p_x = P_s + \delta p_x$. In these equations, capital letters with the subscript '$s$' denote the steady-state (fixed-point) values, while the $\delta$ terms represent small quantum fluctuations around those operating points. For the purpose of evaluating non-classical correlations, such as entanglement, only the fluctuation components need to be considered, as they capture the essential quantum behavior of the system. Nevertheless, it is necessary first to derive the linearized equations of motion under steady-state conditions. This linearization typically assumes certain conditions to ensure system stability, such as $\operatorname{Re}\left\{\mathrm{A}_{s}\right\} \gg 1$ and $|C_s| \gg 1$. In the following analysis, we focus exclusively on the quantum fluctuations and solve the corresponding linearized equations. The quantum fluctuations of the cavity modes are thus analyzed in the vicinity of their respective steady-state values ($A_s$,$P_s$,$X_s$, and $C_s$) as:

\begin{widetext}
    \begin{equation}
    \begin{array}{l}
    \dot{\delta a_{s}}=-\left(j \Delta_{c}+\kappa_{c}\right) \delta \hat{a}_{c}-j g_{o p}\left\{A_{s} \delta \hat{p}_{x} + \hat{p}_{s} \delta \hat{a}_{c}\right\}+\sqrt{2 \kappa_{c}} \delta \hat{a}_{i n} \\ [1em]
    \dot{\delta c_{\omega}}=-\left(j \Delta_{\omega}+\kappa_{\omega}\right) \delta \hat{c}_{\omega}+j g_{\omega p}\left\{\delta \hat{q}_{x} C_{s}+X_{s} \delta \hat{c}_{\omega}\right\}+\sqrt{2 \kappa_{c s}} \delta \hat{c}_{i n} \\ [1em]
    \dot{\delta q_{x}}=\Delta_{e g} \delta \hat{p}_{x}+g_{o p}\left\{\delta \hat{a}_{c}^{\dagger}+\delta \hat{a}_{c}\right\} \\ [1em]
    \dot{\delta p_{x}}=-\gamma_{p} \delta \hat{p}_{x}-\Delta_{e g} \delta \hat{q}_{x}+g_{\omega p}\left\{C_{s} \delta \hat{c}_{\omega}^{\dagger}+C_{s}^{*} \delta \hat{c}_{\omega}\right\}+\delta \hat{b}_{i n}
\end{array}
\label{eq:q_optical_cavity5}
\end{equation}
\end{widetext}

The CV entanglement can arise from the interaction between cavity modes \cite{salmanogli2019entanglement, barzanjeh2011entangling, aggarwal2014selective, vitali2007optomechanical}. In order to analyze the entanglement specifically between the OC and MC, it is essential to define the corresponding quadrature fluctuations, which serve as the primary modes of interest. The matrix representation of these quadrature operators is given by:

\begin{widetext}
\begin{equation}
\resizebox{\textwidth}{!}{$
\begin{bmatrix}
\dot{\delta q_{x}} \\
\dot{\delta p_{x}} \\
\dot{\delta X_{c}} \\
\dot{\delta Y_{c}} \\
\dot{\delta X_{\omega}} \\
\dot{\delta Y_{\omega}}
\end{bmatrix}
=
\underbrace{
\begin{bmatrix}
0 & \Delta_{eg} & \sqrt{2} g_{op} & 0 & 0 & 0 \\
-\Delta_{eg} & -\gamma_{p} & 0 & 0 & \sqrt{2} g_{\omega p} C_{sr} & -\sqrt{2} g_{\omega p} C_{si} \\
0 & \sqrt{2} g_{op} A_{si} & -\kappa_{c} & \Delta_{c} + g_{op} P_{s} & 0 & 0 \\
0 & -\sqrt{2} g_{op} A_{sr} & -\Delta_{c} - g_{op} P_{s} & -\kappa_{c} & 0 & 0 \\
-\sqrt{2} g_{\omega p} C_{si} & 0 & 0 & 0 & -\kappa_{\omega} & \Delta_{\omega} - g_{wp} q_{s} \\
\sqrt{2} g_{wp} C_{sr} & 0 & 0 & 0 & -\Delta_{\omega} - g_{wp} q_{s} & -\kappa_{\omega}
\end{bmatrix}
}_{A_{ij}}
\underbrace{
\begin{bmatrix}
\delta q_{x} \\
\delta p_{x} \\
\delta X_{c} \\
\delta Y_{c} \\
\delta X_{\omega} \\
\delta Y_{\omega}
\end{bmatrix}
}_{u(0)}
+
\underbrace{
\begin{bmatrix}
0 \\
\delta b_{in} \\
\sqrt{2\kappa_{c}} \, \delta X_{c}^{in} \\
\sqrt{2\kappa_{c}} \, \delta Y_{c}^{in} \\
\sqrt{2\kappa_{\omega}} \, \delta X_{\omega}^{in} \\
\sqrt{2\kappa_{\omega}} \, \delta Y_{\omega}^{in}
\end{bmatrix}
}_{n(t)}
$}
\label{eq:q_optical_cavity6}
\end{equation}
\end{widetext}

In Eq. \ref{eq:q_optical_cavity6}, $\delta X_c^{in}$, $\delta Y_c^{in}$, $\delta X_{\omega}^{in}$, and $\delta Y_{\omega}^{in}$ are the noise quadrature operator. The solution of Eq. \ref{eq:q_optical_cavity6} produces an equation with a form as "$u(t) = e^{(A_{i,j})}u(0) +\int(e^{(A_{i,j}s)}n(t-s))ds$", where $n(s)$ is the noise. For this equation, the input noises obey the correlation function \cite{salmanogli2019entanglement, barzanjeh2011entangling, aggarwal2014selective, vitali2007optomechanical}. The entanglement between the cavity modes can be quantitatively analyzed using the Symplectic eigenvalue criterion, as described in \cite{lauk2020perspectives, simon2000peres}:

\begin{widetext}
\begin{equation}
    \eta=\frac{1}{\sqrt{2}} \sqrt{\sigma \pm \sqrt{\sigma^{2}-2 \operatorname{det}(\sigma)}}, \sigma=\operatorname{det}(A)+\operatorname{det}(B)-2 \operatorname{det}(C)
    \label{eq:q_optical_cavity7}
\end{equation}
\end{widetext}

Eq. \ref{eq:q_optical_cavity7} justifies that "$2 \eta >= 1$" stands for separability of modes; otherwise, two modes become entangled \cite{simon2000peres}.

\begin{figure*}[htbp]
    \centering
    % Row 1: (a) and (b)
    \begin{subfigure}[b]{0.5\textwidth}
        \centering
        \includegraphics[width=\linewidth]{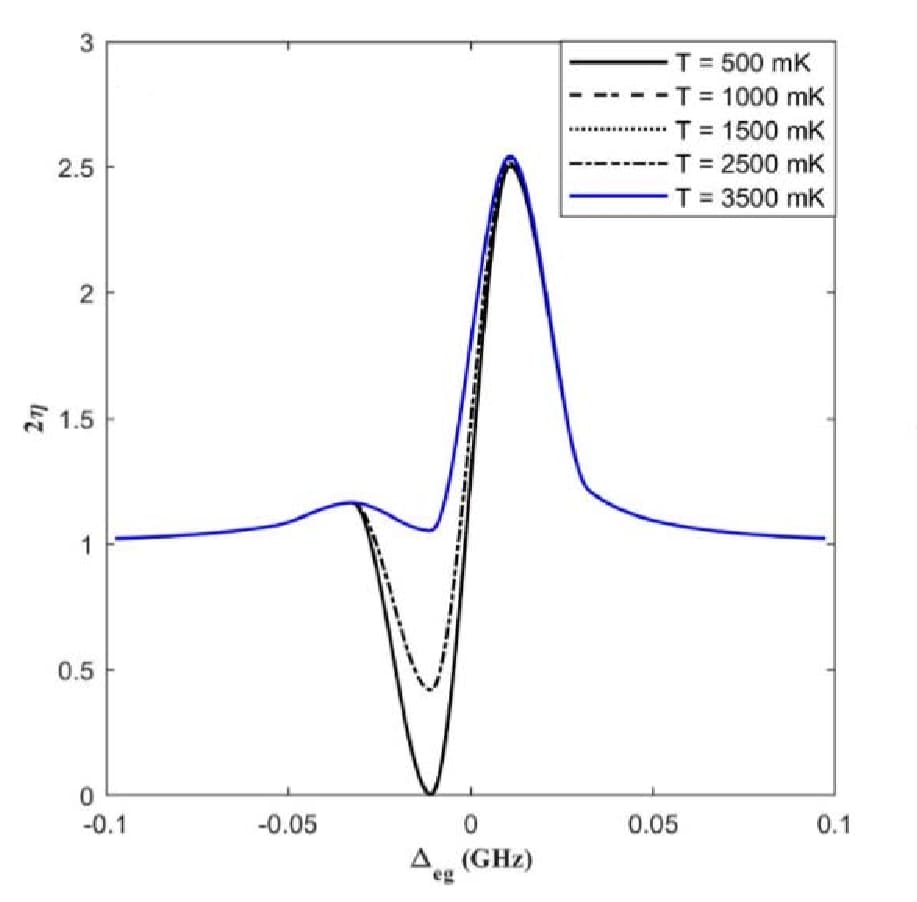}
        \caption{}
        \label{fig:q_temperature_effect_a}
    \end{subfigure}
    \hfill
    \begin{subfigure}[b]{0.48\textwidth}
        \centering
        \includegraphics[width=\linewidth]{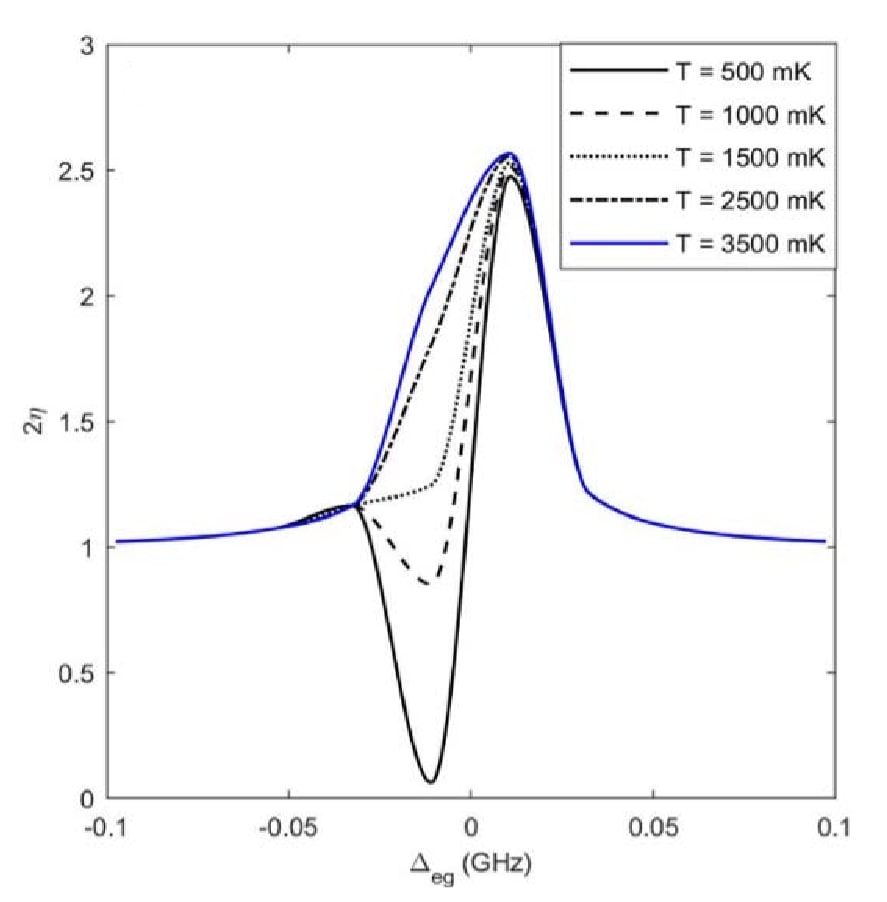}
        \caption{}
        \label{fig:q_temperature_effect_b}
    \end{subfigure}

        \caption{Temperature effect on entanglement between: (a) $a_c$ and $c_{\omega}$, (b) $a_c$ and $c_b$ for $\mu_c = 0.0002$ and $D_{td} = 20$ m $\kappa_{atm} = 2 \times10^{-6}$ 1/m, and $\kappa_t = 18.2$ 1/m \cite{salmanogli2021entanglement}.}
    \label{fig:composite3}
\end{figure*}

Two-mode entanglement between the mode $a_c$ and $c_{\omega}$ as well as between $a_c$ and $c_b$ is analyzed under the conditions of $\mu_c = 0.0002$ and a transmitter-detector distance $D_{td} = 20$ m. In this context (see Fig. \ref{fig:q_optoelectronic_system_schematic_a}, $a_c$ represents the OC mode, $c_w$ denotes the MC mode, and $c_b$ corresponds to the backscattered mode of the target prior to detection. The results are illustrated in Fig. \ref{fig:composite3}, where the entanglement measure $2 \eta$ is plotted as a function of the PD detuning frequency. A peak in entanglement is observed near $\Delta_{eg} \approx 0$, indicating that the PD is optimally excited when its frequency $\omega_c$ is nearly resonant with the transition frequency $\omega_{eg}$. As shown in Fig. \ref{fig:q_temperature_effect_a}, significant entanglement persists even when the cavity temperature $T_c$ reaches 3500 mK. This finding challenges a well-known limitation of traditional tripartite systems, which are typically used for generating entangled photons: at elevated temperatures, thermally excited photons degrade entanglement. However, once the MC-generated photons propagate through the atmosphere toward the target, the signal becomes unpredictable due to environmental effects and target-induced scattering. These influences are modeled using Eq. \ref{eq:q_optical_cavity7}, with the results shown in Fig. \ref{fig:q_temperature_effect_b}. It becomes evident that for $T_c > 1000 mK$, the returned backscattered signals become completely separable, and entanglement vanishes. Therefore, particular attention must be given to the subsystem responsible for generating entangled photons. Although the effects of atmospheric propagation and target scattering cannot be controlled, a carefully designed subsystem can mitigate their impact \cite{salmanogli2021entanglement}. By enabling entangled-state generation at elevated temperatures, such a design helps counteract the detrimental effects of environmental noise and preserve the quantum correlations essential for detection and communication tasks.

\begin{figure*}
    \centering
    \includegraphics[width=0.75\textwidth]{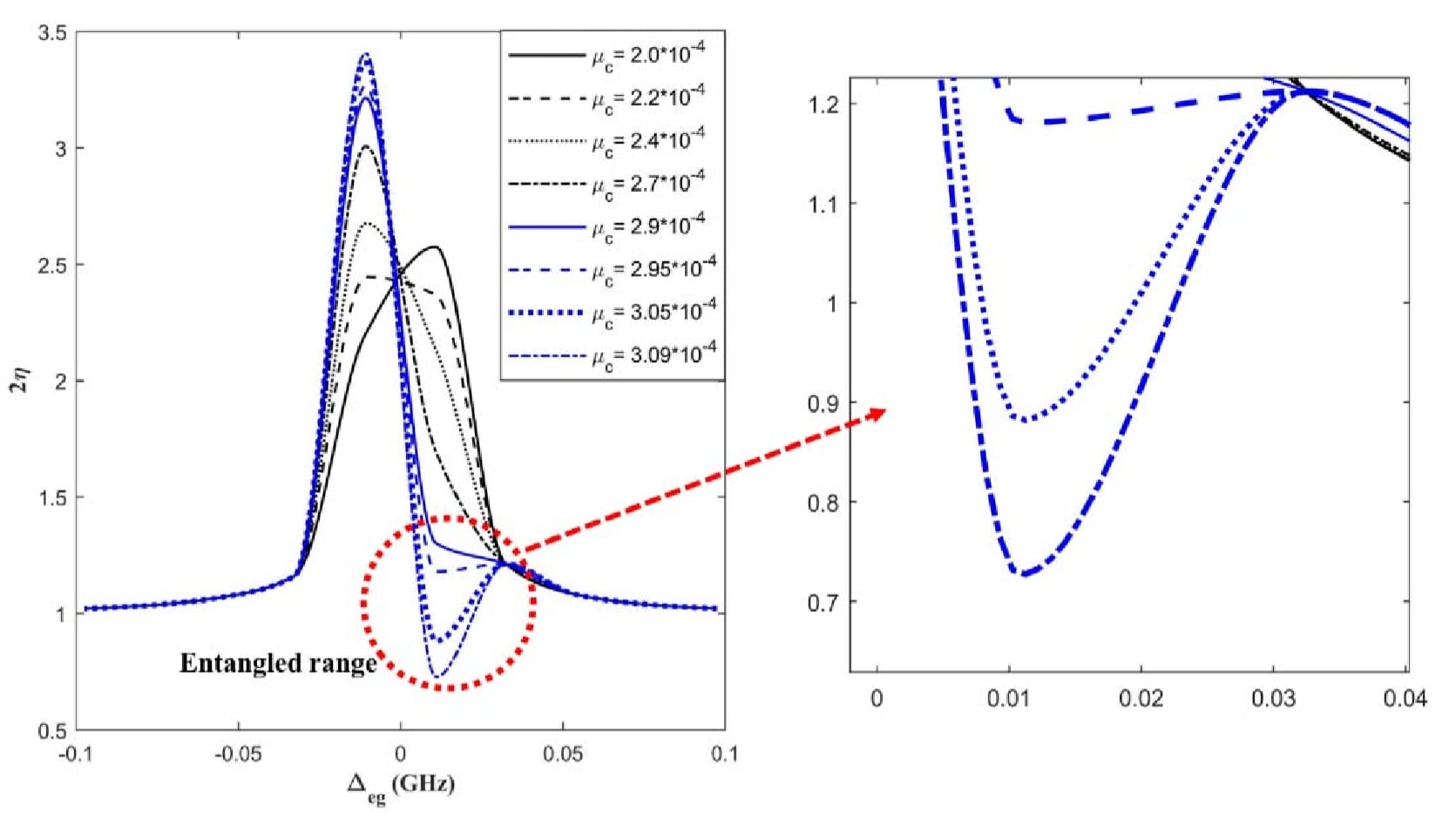}
    \caption{\label{fig:q_MC-PD_coupling} MC-PD coupling effect on entanglement between $a_c$ and $c_{\omega}$ at $T_c = 5000 mK$ and $D_{td} = 20$ m \cite{salmanogli2021entanglement}.}
\end{figure*}

Another important factor here is the sensitivity of the variable capacitance to the electron-hole displacement labeled with $\mu_c$. It is clearly evident in Fig. \ref{fig:q_MC-PD_coupling} that increasing the coupling strength between MC and PD leads to a substantial enhancement in the entanglement between the quantum states. This coupling plays a pivotal role in distinguishing the present approach, which relies on conventional tripartite systems where such coupling was either minimal or not actively optimized. To better highlight this behavior, a zoomed-in view of the critical region is provided in the inset of the Fig. \ref{fig:q_MC-PD_coupling}. The enhanced MC–PD coupling not only strengthens quantum correlations but also facilitates the preservation of entangled states at elevated operational temperatures. This capability marks a significant step toward practical quantum sensing and communication in real-world environments. Specifically, the stronger coupling helps to overcome the detrimental effects of thermal noise that typically degrade entanglement in conventional setups. This makes the system \cite{salmanogli2021entanglement} highly suitable for applications in quantum radar, secure communication, and remote sensing, where environmental control is limited and maintaining quantum coherence is critical.

\begin{figure*}
    \centering
    \includegraphics[width=0.5\textwidth]{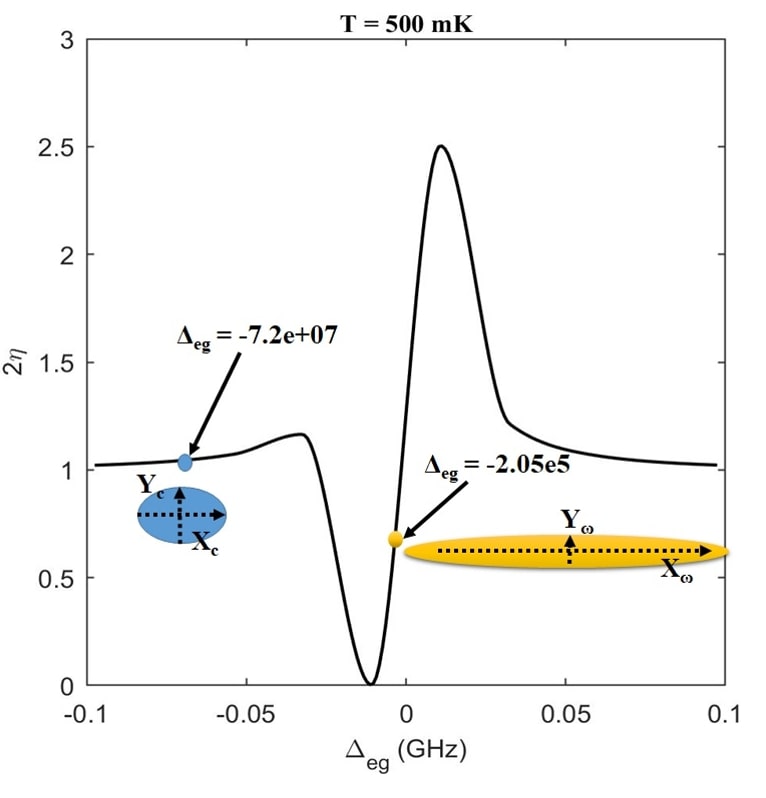}
    \caption{\label{fig:q_entanglement_measure_eta} Entanglement measure $2 \eta$ versus detuning frequency $\Delta_{eg}$ at a fixed temperature of 500 mK. A pronounced peak in entanglement is observed near $\Delta_{eg} = −2.05 \times 105$ Hz, corresponding to strong quantum correlations between cavity modes. Inset ellipses show quadrature phase-space representations at selected detuning points: the yellow ellipse (right) indicates highly squeezed, entangled states, while the blue ellipse (left) represents weaker entanglement with reduced squeezing. These findings emphasize the critical role of detuning in tuning and preserving entanglement.}
\end{figure*}

The nonclassicality generated (squeezing effect) due to the quantum system presented is another important effect that one needs to consider. Fig. \ref{fig:q_entanglement_measure_eta} illustrates the variation of the entanglement measure $2\eta$ as a function of the detuning frequency $\Delta_{eg}$, at an operational temperature of 500 mK. Two distinct entanglement regimes are highlighted. Near $\Delta_{eg} = -2.05 \times 105$ Hz, a sharp peak in $2 \eta$ is observed, indicating a strong entangled state when PD is nearly resonant with the atomic transition frequency $\omega_{eg}$. This strong entanglement is attributed to optimal coupling between the MC and PD modes, enabling coherent quantum correlations. The corresponding quadrature distribution (marked in yellow) shows a well-elongated elliptical shape along the $X_{\omega}$ axis, suggesting strong phase-sensitive correlations. In contrast, at $\Delta_{eg} = -7.2 \times 107$ Hz, the system exhibits a much weaker entanglement, with $2 \eta \approx 1$, which signifies nearly separable states. The corresponding quadrature distribution (marked in blue) appears more circular and localized, indicating reduced squeezing and weaker correlations. These results confirm that optimal detuning is critical to maximize the degree of entanglement and highlight the system’s sensitivity to coupling-induced frequency alignment. This tunability could be exploited in practical implementations to maintain entanglement under varying operational or environmental conditions.

Up to this point, the dynamics of the quantum radar system incorporating the opto-electronic converter have been analytically derived using the canonical quantization method and Heisenberg–Langevin equations. The Symplectic eigenvalue criterion has been applied as a robust tool for evaluating the entanglement between cavity modes. Following the entanglement analysis of the converter modes outlined according to the system architecture in Figs. \ref{fig:composite3}, \ref{fig:q_MC-PD_coupling} and \ref{fig:q_entanglement_measure_eta}. A central objective of this system was to demonstrate that the converter shown in Fig. \ref{fig:q_optoelectronic_system_schematic_a} can be designed such that these environmental effects do not completely destroy the entanglement between modes. One notable distinction between the opto-electronic and electro-opto-mechanical converters lies in the treatment of signal amplification. As highlighted in Fig. \ref{fig:q_optoelectronic_system_schematic_a}, this system deliberately omits any amplifier or active medium. This design choice is based on findings from the literature, which indicate that amplifying entangled microwave photons—particularly to increase photon number while preserving quantum correlations—is extremely challenging and often introduces additional noise that can severely degrade entanglement \cite{ge2015conservation}. Therefore, the focus of opto-electronic system was to show that a quantum radar system based on this converter can function effectively without requiring amplification, and more importantly, that it can preserve the entanglement of the photons during atmospheric propagation and prior to detecting the backscattered signal. This highlights a critical advantage of this system architecture in terms of preserving quantum correlations in realistic operational scenarios.

\subsubsection{Design of A Single Josephson Junction JPA to Create Entangled Photons: Quantum Discord}

A single JJ can function as a parametric amplifier by leveraging its intrinsic nonlinearity to enable energy transfer between electromagnetic modes. Fundamentally, a JJ consists of two superconducting electrodes separated by a thin insulating barrier, allowing Cooper pairs to tunnel through via the Josephson effect \cite{planat2020resonant, eichler2014controlling, macklin2015near, planat2019understanding, chirolli2023quartic, bell2012quantum, yan2020engineering, salmanogli2024blochnium}. The junction exhibits non-linear inductance, which is the key ingredient for parametric amplification. When driven by an external pump tone at frequency $f_{pump}$, the non-linear inductance modulates periodically, allowing for energy exchange between different frequency components. This parametric process can lead to degenerate or non-degenerate amplification \cite{planat2020resonant, eichler2014controlling, macklin2015near, planat2019understanding, chirolli2023quartic, bell2012quantum, yan2020engineering, salmanogli2024blochnium}. In degenerate mode, the signal and idler share the same frequency but have a phase relationship that governs the gain. In non-degenerate mode, the idler frequency is different from the signal, leading to greater operational flexibility. This process is governed by the quantum Langevin equation and the input-output formalism, where the interaction Hamiltonian describes the coupling between the pump, signal, and idler fields. The strength of amplification is dictated by the critical current, $I_c$, of the junction, its shunt capacitance, and the external impedance environment \cite{boutin2017effect}. These parameters influence the bandwidth, gain, and stability of the amplifier. The gain mechanism is fundamentally based on parametric conversion, where the pump injects energy into the signal and idler fields while maintaining overall energy conservation. In the following, this review article centers on the theoretical analysis of single-JJ and arrayed-JJ JPAs, as well as emerging JPA architectures, and compares simulation and modeling results with existing findings in the literature.

Firstly, a single JJ JPA is quantum-mechanically analyzed and it attempts to theoretically derive amplifier critical parameters such as gain. The approach begins with the theoretical derivation of the system's Hamiltonian, followed by an analysis of the dynamics using the quantum Langevin equation. The Hamiltonian of the lumped element JPA (that is, a capacitively shunted JJ coupled to a transmission line \cite{salmanogli2018plasmonic}) is expressed as (setting $\hbar = 1$ for the reminder of the article):

\begin{equation}
    H_{\mathrm{JPA}} = \frac{\hat{Q}^2}{2C} - E_J \cos(\hat{\phi}) 
    \longrightarrow \approx \omega_0 \hat{a}^\dagger \hat{a} + \frac{\Lambda}{6} \left( \hat{a}^\dagger + \hat{a} \right)^4
    \label{eq:q_JJ1}
\end{equation}

where $C$, $E_J$, $Q$, and $\phi$ are the capacitance, Josephson energy, charge, and phase across the junction, respectively. The other side of the arrow is obtained by expanding the cosine to the fourth order and introducing annihilation and creation operators ($\hat{a}$, $\hat{a}^{\dagger}$). In this equation, the Kerr coefficient $\Lambda = -\frac{E_c}{2}$ and the bare frequency of the resonator $\omega = \sqrt{8 E_c E_J}$, where $E_c = \frac{e^2}{2C}$ is the charging energy of the Josephson junction. Using these definitions, the Hamiltonian of a single Josephson junction JPA driven by a monochromatic current pump \cite{boutin2017effect, collett1984squeezing} is expressed as:

\begin{equation}
    H_{CP} = H_{JPA} + \varepsilon e^{(-jw_pt)}\hat{a}^{\dagger} + \varepsilon^* e^{(jw_pt)}\hat{a}
    \label{eq:q_JJ2}
\end{equation}

where $\varepsilon$ and $\omega_p$ are the amplitude of the drive pump and its frequency, respectively. To simplify successive algebra, it becomes versatile to eliminate the pump Hamiltonian using a displacement operator such as $a = \alpha + \delta a$, where $\alpha$ and $\delta a$ are the classic field and quantum fluctuation, respectively. In the displaced frame mentioned above, the Hamiltonian of the system becomes:

\begin{equation}
    H_{CP} = \Delta_{0} \, \delta \hat{a}^\dagger \delta \hat{a} 
    + \frac{\lambda_{1}}{2} \, \delta \hat{a}^{\dagger 2} 
    + \frac{\lambda_{1}^*}{2} \, \delta \hat{a}^2 
    + H_{nc}
    \label{eq:q_JJ3}
\end{equation}

In Eq. \ref{eq:q_JJ3} the shifted detuning is 
\(\Delta_{0} = \omega_{0} + 4|\alpha|^{2} \Lambda - \omega_{p}\), 
and the effective parametric pump strength is 
\(\lambda_{1} = 2\alpha^{2} \Lambda\), and also the classical field can be solved using the quantum Langevin equation \cite{salmanogli2018plasmonic}. 
Finally, the last term in the equation, \(H_{nc}\), is the nonlinear correction Hamiltonian arising from the displacement of the Kerr nonlinearity and can be written as:
\begin{equation}
    H_{nc} = \mu_{0} \, \delta \hat{a}^{\dagger 2} \delta \hat{a} 
    + \mu_{0}^{*} \, \delta \hat{a}^\dagger \delta \hat{a}^{2} 
    + \Lambda \, \delta \hat{a}^{\dagger 2} \delta \hat{a}^{2}
    \label{eq:q_JJ4}
\end{equation}

where \(\mu_{0} = 2\alpha^{2} \Lambda\) is the cubic term coefficient. However, to obtain a linear equation, one can ignore \(H_{nc}\) for the small quantum fluctuation, which is valid in the low-gain and low Kerr nonlinearity regime~\cite{salmanogli2018plasmonic}. It is notable to indicate that the output state of the JPA pumped with a monochromatic current, regarding Eq. \ref{eq:q_JJ3} and Eq. \ref{eq:q_JJ4}, is a displaced squeezed state. In the following, it is focused on Eq. \ref{eq:q_JJ4} by ignoring \(H_{nc}\), and using the input-output formula 
\((\delta \hat{a}_{\text{out}} = \sqrt{\kappa} \delta \hat{a} - \delta \hat{a}_{\text{in}})\)~\cite{salmanogli2018plasmonic}, 
attempt to analytically calculate the gain of the system. Generally, the behavior of the intracavity mode may be calculated by the master equation method; still, one can use the quantum Langevin equation for a single-mode cavity as:

\begin{equation}
    \frac{d \, \delta \hat{a}}{dt} 
    = -j \left[ \delta \hat{a}, \left(H_{CP} - H_{nc}\right) \right] 
    - \frac{\kappa}{2} \delta \hat{a} + \sqrt{\kappa} \, \delta \hat{a}_{\text{in}}
    \label{eq:q_JJ5}
\end{equation}

where \(\kappa\) and \(\delta \hat{a}_{\text{in}}\) are the JPA damping constant created due to any mismatching between the JPA and the environment, and the embedded quantum fluctuation of the external field applied to the system, respectively.  Eq.~\ref{eq:q_JJ5} can be expanded for \(\delta \hat{a}\) and its conjugate as:

\begin{equation}
    \frac{d \, \delta \hat{a}}{dt} = -\left( j \Delta_{0} + \frac{\kappa}{2} \right) \delta \hat{a} - j \lambda_{1} \delta \hat{a}^{\dagger} + \sqrt{\kappa} \, \delta \hat{a}_{\text{in}}
    \label{eq:q_JJ6}
\end{equation}
\begin{equation}
    \frac{d \, \delta \hat{a}^{\dagger}}{dt} = -\left( -j \Delta_{0} + \frac{\kappa}{2} \right) \delta \hat{a}^{\dagger} + j \lambda_{1}^{*} \delta \hat{a} + \sqrt{\kappa} \, \delta \hat{a}_{\text{in}}^{\dagger}
    \label{eq:q_JJ7}
\end{equation}

In Eq. \ref{eq:q_JJ6} and \ref{eq:q_JJ7} are linear and, in terms of frequency components \(\left(a(\omega) = \frac{1}{\sqrt{2\pi}} \int a(t) e^{j \omega t} dt\right)\), becomes:

\begin{widetext}
\begin{equation}
    -j\omega \, \delta \hat{a}(\omega) = -\left( j \Delta_{0} + \frac{\kappa}{2} \right) \delta \hat{a}(\omega) - j \lambda_{1} \delta \hat{a}^{\dagger}(-\omega) + \sqrt{\kappa} \, \delta \hat{a}_{\text{in}}(\omega)
    \label{eq:q_JJ8}
\end{equation}
\end{widetext}

\begin{widetext}
\begin{equation}
    j\omega \, \delta \hat{a}^{\dagger}(-\omega) = -\left( -j \Delta_{0} + \frac{\kappa}{2} \right) \delta \hat{a}^{\dagger}(-\omega) + j \lambda_{1}^{*} \delta \hat{a}(\omega) + \sqrt{\kappa} \, \delta \hat{a}_{\text{in}}^{\dagger}(-\omega)
    \label{eq:q_JJ9}
\end{equation}
\end{widetext}

The last step to calculate the gain is to establish the relating scattering matrix, then employing the input-output formula, the gain of the signal and idler can be calculated \cite{eichler2014controlling, salmanogli2023entangled, cha2020inp}. It is shown as follows:

\begin{widetext}
\begin{equation}
\begin{aligned}
&\left\{
\begin{aligned}
&\left[ -j(\omega+\Delta_0) + \frac{\kappa}{2} \right] \delta \hat{a}(\omega) 
+ j \lambda_1 \, \delta \hat{a}^\dagger(-\omega) 
= \sqrt{\kappa} \, \delta \hat{a}_{\mathrm{in}}(\omega), \\
&\left[ j(\omega-\Delta_0) + \frac{\kappa}{2} \right] \delta \hat{a}^\dagger(-\omega) 
- j \lambda_1^* \, \delta \hat{a}(\omega) 
= \sqrt{\kappa} \, \delta \hat{a}^\dagger_{\mathrm{in}}(-\omega)
\end{aligned}
\right. \\[6pt]
&\Rightarrow 
\left[
\begin{matrix}
\delta \hat{a}_{\mathrm{out}}(\omega) \\
\delta \hat{a}^\dagger_{\mathrm{out}}(-\omega)
\end{matrix}
\right]
= 
\left\{
\kappa
\left[
\begin{matrix}
-j(\omega+\Delta_0)+\frac{\kappa}{2} & j\lambda_1 \\
-j\lambda_1^* & j(\omega-\Delta_0)+\frac{\kappa}{2}
\end{matrix}
\right]^{-1}
- \begin{pmatrix}
1 & 0 \\
0 & 1
\end{pmatrix}
\right\}
\left[
\begin{matrix}
\delta \hat{a}_{\mathrm{in}}(\omega) \\
\delta \hat{a}^\dagger_{\mathrm{in}}(-\omega)
\end{matrix}
\right]
\end{aligned}
\label{eq:q_JJ10}
\end{equation}
\end{widetext}

Finally, the gain matrix is presented as:

\vspace{1em} % before the equation
\begin{widetext}
\begin{equation}
\begin{bmatrix}
\delta \hat{a}_{\text{out}}(\omega) \\
\delta \hat{a}^\dagger_{\text{out}}(-\omega)
\end{bmatrix}
=
\begin{bmatrix}
\frac{\kappa \left[ j(\omega - \Delta_0) + \frac{\kappa}{2} \right]}{\frac{\kappa^2}{4} - i\kappa\Delta_0 + \omega^2 - \Delta_0^2} - 1 &
\frac{-j \lambda_1 \kappa}{\frac{\kappa^2}{4} - j\kappa\Delta_0 + \omega^2 - \Delta_0^2} \\[10pt]
\frac{j \lambda_1^* \kappa}{\frac{\kappa^2}{4} - j\kappa\Delta_0 + \omega^2 - \Delta_0^2} &
\frac{\kappa \left[ -i(\omega + \Delta_0) + \frac{\kappa}{2} \right]}{\frac{\kappa^2}{4} - j\kappa\Delta_0 + \omega^2 - \Delta_0^2} - 1
\end{bmatrix}
\begin{bmatrix}
\delta \hat{a}_{\text{in}}(\omega) \\
\delta \hat{a}^\dagger_{\text{in}}(-\omega)
\end{bmatrix}
\label{eq:q_JJ11}
\end{equation}
\end{widetext}
\vspace{1em} % before the equation

where the matrix element $(1,1)$ determines the signal gain and the element $(1,2)$ signifies the idler gain. It should be noted that the effective parametric pump strength affects the idler gain, not the signal gain; in contrast, the signal gain depends on the decay of the quantum system and the frequency detuning. The gain of the single JJ JPA is defined as:

\begin{widetext}
\begin{equation}
G_{\mathrm{single\_JJ\_JPA}} =
\left\{
\frac{
\kappa \left[ j(\omega - \Delta_0) + \frac{\kappa}{2} \right]
}{
\frac{\kappa^2}{4} - j \kappa \Delta_0 + \omega^2 - \Delta_0^2
} - 1
\right\}
-
\left\{
\frac{
- j \lambda \kappa
}{
\frac{\kappa^2}{4} - j \kappa \Delta_0 + \omega^2 - \Delta_0^2
}
\right\}
\label{eq:q_JJ12}
\end{equation}
\end{widetext}

where the first part signifies the signal gain and the second part determines the idler gain. Fortunately, a compact formula can be presented for two-mode squeezed thermal states (zero-mean Gaussian states) to reduce the covariance matrix (CM) into the standard form \cite{serafini2023quantum, weedbrook2012gaussian, adesso2007entanglement}. Consequently, the CM of the selected modes in the system can be presented in the form of the following matrix:

\begin{equation}
V_{AB} =
\begin{pmatrix}
(\tau b + \eta) \mathbf{I} & \sqrt{\tau (b^2 - 1)} \, \mathbf{C} \\
\sqrt{\tau (b^2 - 1)} \, \mathbf{C} & b \mathbf{I}
\end{pmatrix}
\label{eq:VAB}
\end{equation}

where $ \mathbf{I} \equiv \mathrm{diag}(1,1)$, $\mathbf{C} \equiv \mathrm{diag}(1,-1)$,
$a = n_{o1} + 0.5$, $b = n_{o2} + 0.5$,
$\tau = d_{o12}^2 / (b^2 - 1)$, $\eta = a - (b_2^* \, d_{o12}^2 / (b^2 - 1))$ \cite{serafini2023quantum, weedbrook2012gaussian, adesso2007entanglement}.

In these equations, $a$ and $b$ are the expectation values of the I/Q signals for two oscillators, derived as $
a = \langle I_1(\omega) I_1(\omega) \rangle = \langle Q_1(\omega) Q_1(\omega) \rangle, \quad
b = \langle I_2(\omega) I_2(\omega) \rangle = \langle Q_2(\omega) Q_2(\omega) \rangle$,
where $
I = \frac{\delta \hat{a}_i^\dagger + \delta \hat{a}_i}{\sqrt{2}}, \quad
Q = \frac{\delta \hat{a}_i - \delta \hat{a}_i^\dagger}{j\sqrt{2}}, \quad i=1,2$.

The parameters $n_{o1}$, $n_{o2}$, and $d_{o12}$ are the output mean photon numbers of the first and second oscillator, and the cross-correlation (phase-sensitive), respectively.  
One can use the input-output formula \cite{salmanogli2020entanglement} to calculate the output mean photon numbers, 
$n_{o1} = 2\kappa_1 \langle \delta a_1^\dagger \delta a_1 \rangle + \langle \delta a_{\text{in-1}}^\dagger \delta a_{\text{in-1}} \rangle, \quad
n_{o2} = 2\kappa_2 \langle \delta a_2^\dagger \delta a_2 \rangle + \langle \delta a_{\text{in-2}}^\dagger \delta a_{\text{in-2}} \rangle$, $d_{o12} = 2\sqrt{\kappa_1 \kappa_2} \langle \delta a_1 \delta a_2 \rangle$.Finally, the mean photon numbers of the oscillators are 
$n_1 \equiv \langle \delta a_1^\dagger \delta a_1 \rangle, \quad
n_2 \equiv \langle \delta a_2^\dagger \delta a_2 \rangle, \quad
d_{12} \equiv \langle \delta a_1 \delta a_2 \rangle$,
where $n_1$, $n_2$, and $d_{12}$ represent the first oscillator mean photon number, the second oscillator mean photon number, and the cross-correlation between the two oscillators, respectively.  
It is assumed that $d_{12}$ is real-valued.

The output entropy associated with heterodyne detection is equal to the average entropy of the output ensemble $A$ \cite{serafini2023quantum, weedbrook2012gaussian, adesso2007entanglement}.  
Since entropy is invariant under displacements, it can be written as$ S(A_{\mathrm{hetero}}) = h(\tau + \eta)$,
where $h(x) \equiv (x+0.5) \log_2 (x+0.5) - (x-0.5) \log_2 (x-0.5)$. The von Neumann entropy of an $n$-mode Gaussian state with CM $V_{AB}$ is $S(V_{AB}) = \sum_{i=1}^{N} f(\nu_i)$,
where $\nu_i$ are the symplectic eigenvalues. However, there is a heterodyne detection for which it is optimal for minimization of the output entropy, so the Gaussian discord is optimal. The “Gaussian quantum discord” of a two-mode Gaussian state $\rho_{AB}$, assuming a two-mode squeezed thermal state in this study, can be defined as the quantum discord satisfying the conditional entropy restricted to the generalized Gaussian POVMs on B \cite{ge2015conservation}.

Finally, the compact form of quantum discord, classical correlation, and quantum mutual information are given, respectively, by 
$D(\rho_{AB}) = h(b) - h(\nu_-) - h(\nu_+) + h(\tau + \eta)$,
$C(\rho_{AB}) = h(a) - h(\tau + \eta)$,
$I(\rho_{AB}) = h(a) - h(\nu_-) - h(\nu_+)$,
where $\nu_{\pm}$ is the Symplectic eigenvalue of the CM. The Symplectic eigenvalues are defined as  
$\nu_{\pm} = \frac{\Delta \pm \sqrt{\Delta^2 - 4D}}{2}$, where  
$\Delta = \det(a\mathbf{I}) + \det(b\mathbf{I}) + 2\det(d_{o12}\mathbf{I})$, and  $D = \det(V_{AB})$;
in the recent formulas $\det\{\cdot\}$ stands for the matrix determinant.

In the compact form of the equation defined for the quantum discord, the first term stands for the von Neumann entropy of the second oscillator in the system. The second and third terms define the von Neumann conditional entropy of the system. The last term in the equation is the effect of the classical correlation depending on the type of measurement performed on the second oscillator. As a result, in the system defined, the second oscillator entropy and the off-diagonal elements in the CM significantly affect the system's quantum discord. Therefore, particular attention is paid to the interference between oscillators, and an analysis is conducted to determine which quantities may affect the quantum discord. Perhaps the other critical factor that may be considered to enhance the quantum discord is $h(\tau + \eta)$, by which the classical correlation is decreased. As mentioned in \cite{serafini2023quantum, weedbrook2012gaussian, adesso2007entanglement}, this critical factor strongly depends on the type of measurements.

\begin{figure*}
    \centering
    \includegraphics[width=0.75\textwidth]{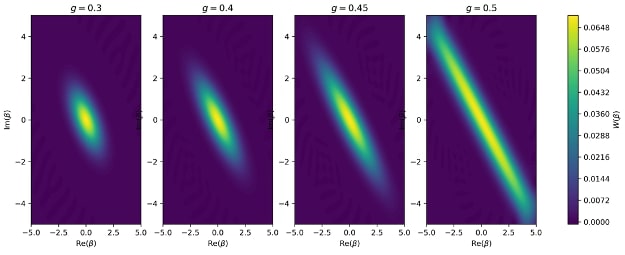}
    \caption{\label{fig:q_squeezing} Steady-state Wigner functions of a nonlinear JPA for varying parametric gain factors $g$. The Wigner function $W(\beta)$ is plotted in the complex phase space, showing increasing squeezing and nonclassicality as $g$ increases from $0.3$ to $0.5$. The results highlight the transition of the JPA output toward a highly squeezed quantum state, suitable for entangled photon generation.}
\end{figure*}

The Wigner function, $W(\beta)$, visualized over the complex phase-space coordinates $\mathrm{Re}(\beta)$ and $\mathrm{Im}(\beta)$, illustrates how the quantum state evolves as $g$ increases from $0.3$ to $0.5$ in Fig. \ref{fig:q_squeezing}. For low $g$, the state remains relatively isotropic, indicating limited squeezing. As $g$ increases, the Wigner function becomes more elongated and tilted, indicating stronger phase-sensitive squeezing and the emergence of nonclassical features---hallmarks of quantum entanglement potential. Notably, at $g=0.5$, the state becomes highly squeezed along one axis, confirming the amplifier’s ability to generate highly correlated photon pairs. This figure confirms that by adjusting the parametric interaction strength, the JPA can transition from a thermal-like state to a nonclassical squeezed state, a crucial feature for quantum information processing and entangled photon generation.

\section{Quantum Radar Subsystems} \label{sec3}

\subsection{Quantum Transmitter}

In quantum radar systems, the quantum transducer plays a pivotal role in enabling coherent and efficient interfacing between the microwave and optical domains \cite{barzanjeh2015microwave}. This is essential in architectures where entangled microwave signals - used to detect a target - must be stored, converted, or measured via optical systems due to their superior detection efficiency and transmission properties. 

In this context, a quantum transducer is a device capable of bidirectional and phase-coherent conversion between microwave and optical photons, ideally without adding thermal noise or decoherence. The fundamental mechanism underpinning this conversion is the tripartite interaction among microwave, mechanical, and optical modes, a framework known as electro-optomechanics or opto-electro-mechanics.

In a typical implementation, the EOM transducer consists of a mechanical resonator that simultaneously couples to both a superconducting microwave cavity and an optical Fabry–Pérot cavity. This configuration enables the mechanical mode to mediate energy exchange between the microwave and optical domains. When both cavities are pumped appropriately, the system realizes an effective beam-splitter-type or two-mode squeezing interaction, depending on detuning configurations, enabling quantum state transfer or entanglement generation across the frequency gap.

\begin{figure*}
    \centering
    \includegraphics[width=0.75\textwidth]{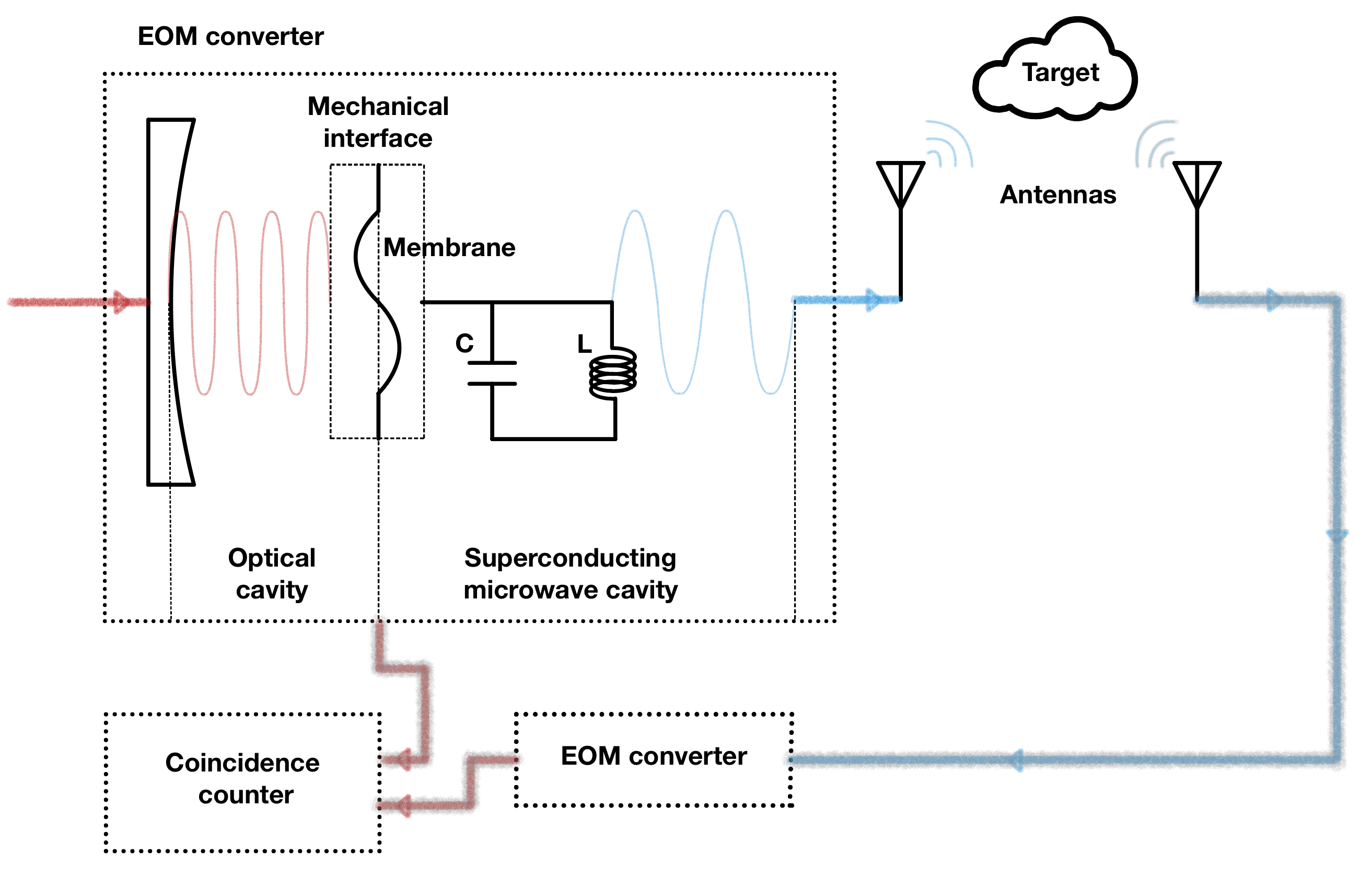}
    \caption{\label{fig:q_transducer} Quantum radar architecture using EOM transduction for microwave-optical signal conversion \cite{barzanjeh2015microwave}.}
\end{figure*}

Figure \ref{fig:q_transducer} illustrates a simplified version of the proposed quantum radar architecture based on EOM transduction proposed in \cite{barzanjeh2015microwave}, which enables coherent photon conversion between the optical and microwave domains. An entangled photon source injects optical signals into a high-fidelity optical cavity, which is mechanically coupled to a superconducting microwave cavity via a suspended membrane. This tripartite system—comprising optical, mechanical, and microwave modes—functions as a quantum transducer, converting optical photons into microwave photons via radiation-pressure-mediated interactions. The generated microwave signal is transmitted toward a distant target via antenna, where it scatters and partially reflects back. In actual implementation, there is no free-space propagation; the antenna is included in the illustration only to aid conceptual understanding \cite{barzanjeh2015microwave}. Upon reception, the microwave echo is routed into a second EOM converter, where it is upconverted to the optical domain. This returned optical signal is jointly measured with the stored idler photon using a coincidence counter, enabling quantum correlation analysis. By preserving non-classical correlations between the transmitted and received modes, this system enhances detection sensitivity in noisy environments, exemplifying the potential of hybrid transduction-based quantum radar \cite{barzanjeh2015microwave}.

\begin{figure*}
    \centering
    \includegraphics[width=0.75\textwidth]{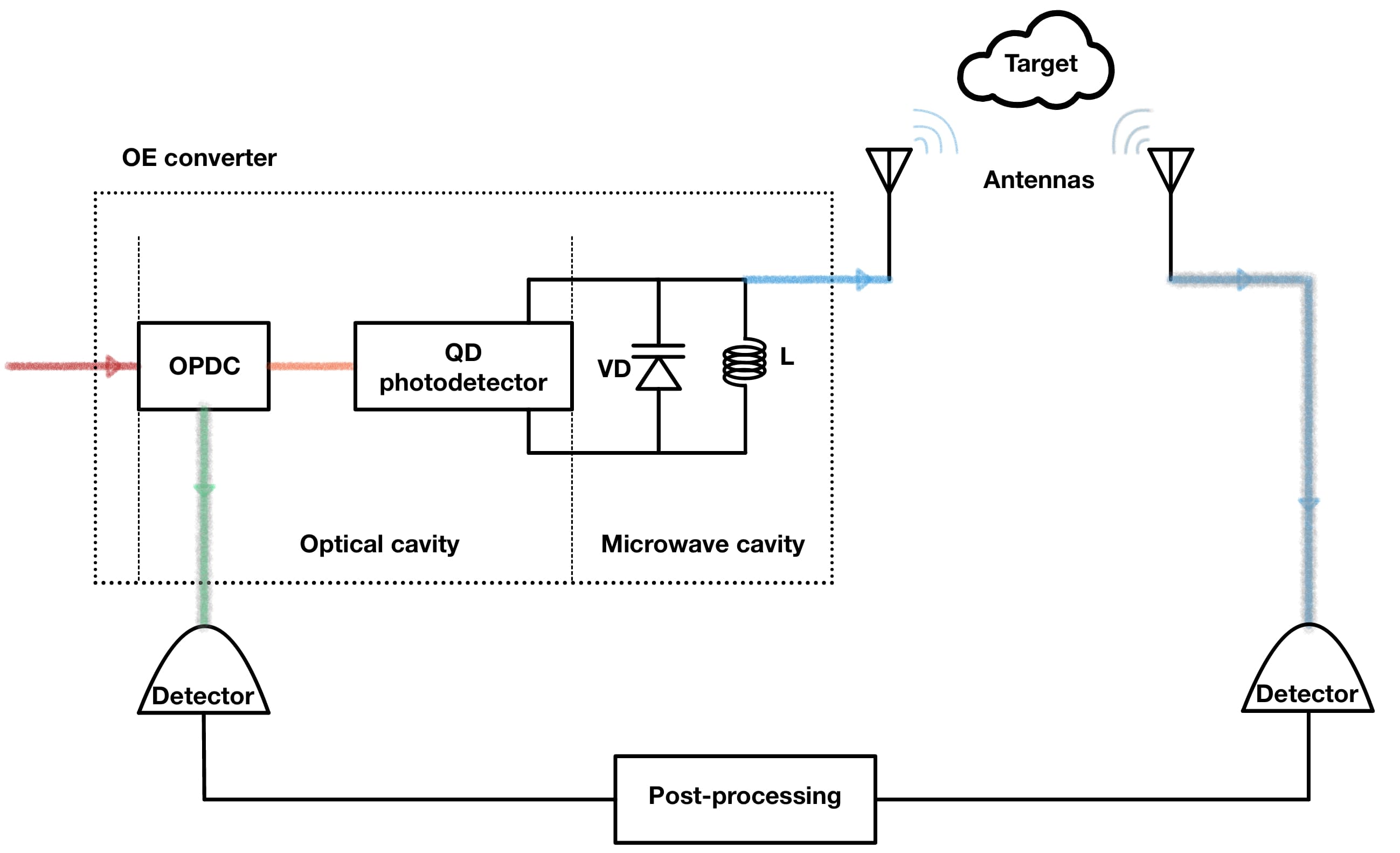}
    \caption{\label{fig:q_transducer_oe} Opto-electronic quantum radar with optical idler storage and microwave signal transmission \cite{salmanogli2021entanglement}.}
\end{figure*}

The opto-electronic (OE) quantum radar illustrated in Fig.\ref{fig:q_transducer_oe} employs a hybrid interface where the entangled photon pairs generated in the optical parametric downconverter (OPDC) are split into a signal and an idler. The idler photon remains in the optical domain and is stored locally as a reference, while the signal photon excites a quantum-dot photodetector. The resulting photocurrent modulates the VD, which tunes the resonance of a microwave cavity to generate microwave photons. These photons are transmitted toward the target and, after scattering and returning through a noisy channel, are detected alongside the retained idler. The detection process relies on the statistical correlation between the optical idler and the returned microwave mode, where the presence of nonclassical correlations indicates the presence of the target.

Josephson-based parametric devices are central to the implementation of microwave quantum radar systems, providing the essential functionality for entanglement generation, amplification, and frequency conversion in the cryogenic domain \cite{chang2019quantum, barzanjeh2020microwave, livreri2022microwave}. Among these, the JPA has been extensively used as a narrowband quantum-limited amplifier capable of generating two-mode squeezed vacuum (TMSV) states by modulating a nonlinear inductance based on the Josephson effect. JPAs are typically used in early stage quantum radar prototypes, where they serve as an entangled signal-idler source in quantum illumination or quantum-enhanced noise radar schemes \cite{chang2019quantum, luong2019receiver}. Despite their limited bandwidth (on the order of a few MHz), JPAs offer excellent noise performance and phase-sensitive gain, making them suitable for proof-of-principle demonstrations of quantum correlations at microwave frequencies \cite{chang2019quantum, luong2019receiver}. 

In addition to JPA, the JPC is designed to perform both non-degenerate amplification and frequency conversion between two orthogonal microwave modes, typically referred to as signal and idler \cite{barzanjeh2020microwave}. JPCs operate via three-wave mixing and exhibit low added noise, making them especially useful in experiments that require simultaneous amplification and separation of entangled modes. They have been used in quantum illumination implementations that adopt digital receivers, where signal and idler beams are downconverted and processed electronically to extract second-order correlations \cite{barzanjeh2020microwave}. However, similar to JPAs, JPCs suffer from limited dynamic range and bandwidth.

To overcome these limitations, the JTWPA has emerged as a broadband high-gain entangled microwave source with minimal added noise \cite{macklin2015near}. Unlike cavity-based JPAs and JPCs, the JTWPA supports traveling wave propagation of microwave signals along a dispersion-engineered nonlinear transmission line, enabling squeezing and amplification over GHz-scale bandwidths. Unlike cavity-based JPAs and JPCs, the JTWPA supports traveling wave propagation of microwave signals along a dispersion-engineered nonlinear transmission line, enabling squeezing and amplification over GHz scale bandwidths \cite{macklin2015near}. This makes JTWPAs particularly well suited for scalable and long-range quantum radar architectures. In recent work, JTWPAs have been employed in a radar prototype implementing quantum illumination with a phase-conjugate receiver, allowing for enhanced detection performance even in the presence of amplifier-induced decoherence and thermal noise \cite{livreri2022microwave}.

\begin{figure}
    \centering
    \includegraphics[width=0.25\textwidth]{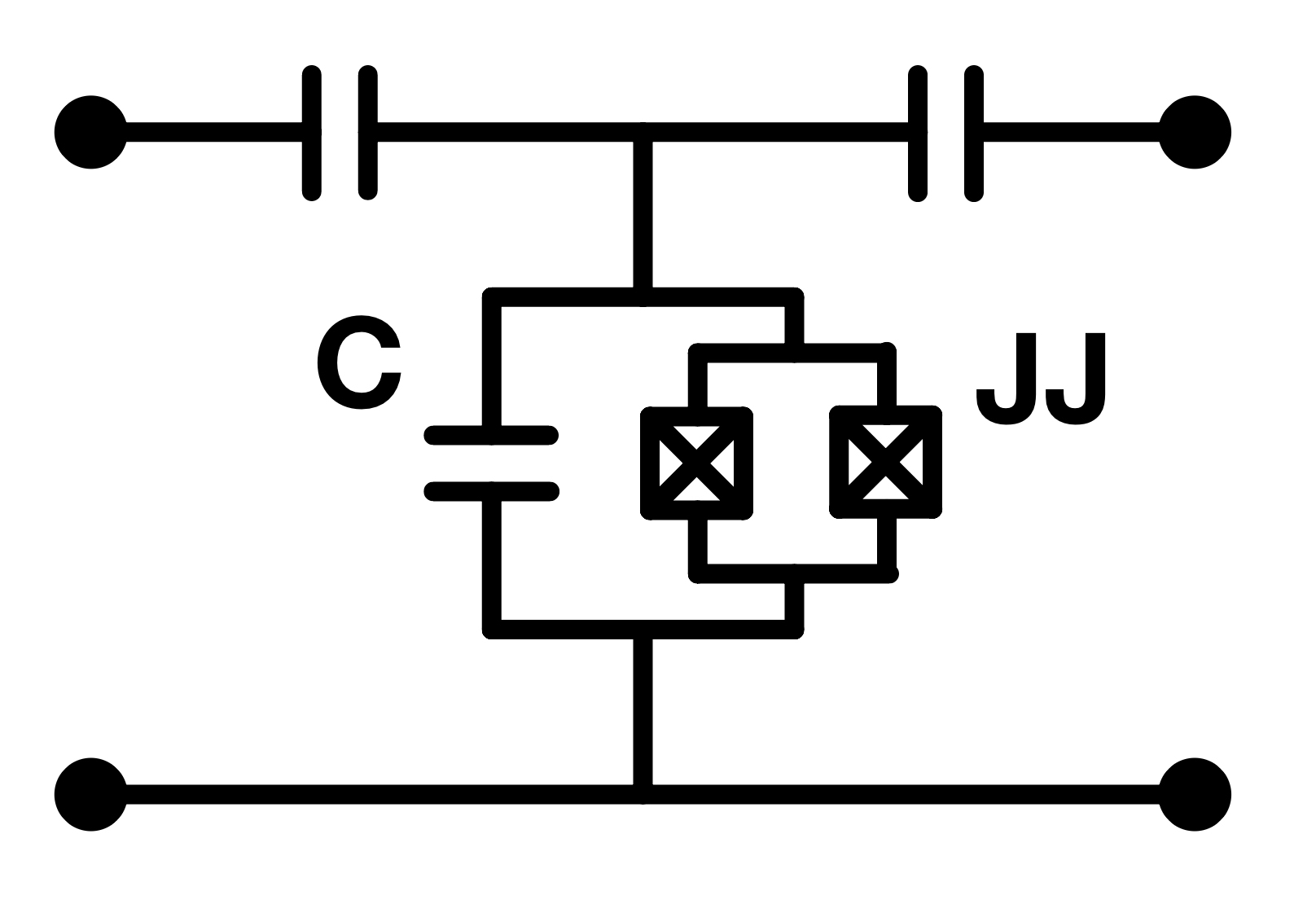}
    \caption{\label{fig:q_JPA} The JPA consisting of a JJ or a tunable SQUID shunted to a lumped-element LC resonator \cite{yurke1989observation, chang2019quantum}.}
\end{figure}

\begin{figure*}
    \centering
    \includegraphics[width=0.5\textwidth]{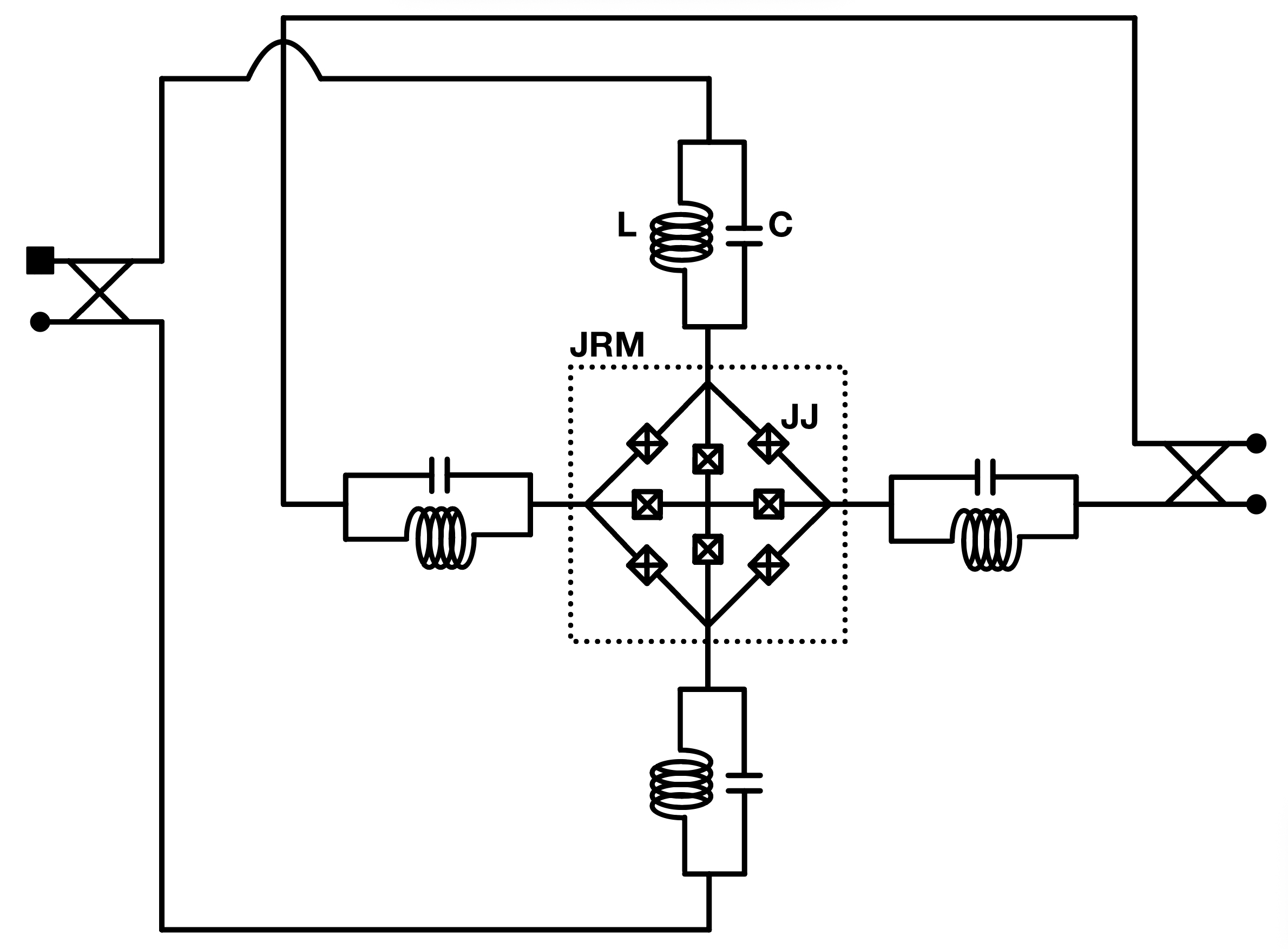}
    \caption{\label{fig:q_JPC} The JPC based on a JRM composed of four Josephson junctions arranged in a Wheatstone bridge configuration \cite{bergeal2010phase, barzanjeh2015microwave}.}
\end{figure*}

\begin{figure*}
    \centering
    \includegraphics[width=0.75\textwidth]{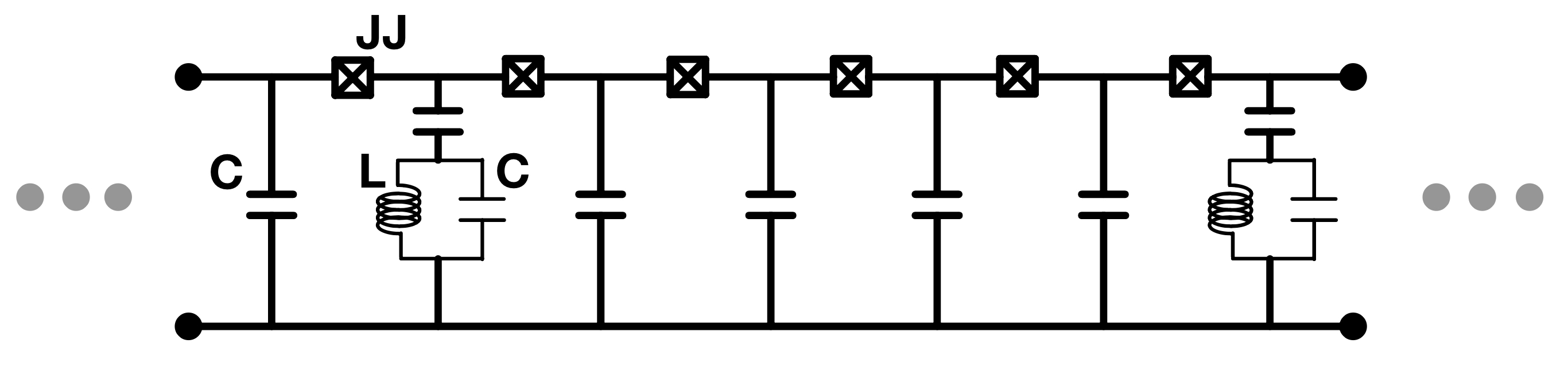}
    \caption{\label{fig:q_JTWPA} The JTWPA composed of a cascaded array of unit cells containing Josephson junctions and capacitive elements \cite{macklin2015near, livreri2022microwave, qiu2023broadband}.}
\end{figure*}

The three circuit schematics illustrate the fundamental designs of Josephson-based parametric devices. Fig. \ref{fig:q_JPA}, depicts the JPA, which consists of a lumped element LC resonator shunted by a Josephson junction (JJ) or a tunable SQUID \cite{yurke1989observation, chang2019quantum}. Fig. \ref{fig:q_JPC} shows the JPC, based on a Josephson Ring Modulator (JRM) composed of four Josephson junctions arranged in a Wheatstone bridge configuration \cite{bergeal2010phase, barzanjeh2015microwave}. Fig. \ref{fig:q_JTWPA} presents the architecture of the JTWPA, which uses a long, dispersion-engineered nonlinear transmission line made of cascaded unit cells containing Josephson junctions and capacitive elements \cite{macklin2015near, livreri2022microwave, qiu2023broadband}.

\subsection{Atmosphere, Channel, and Related Loss}

Atmospheric losses are frequency-dependent. Both radar and lidar systems often operate in frequency bands where atmospheric transmission is relatively low. In general, attenuation and scattering are less severe at microwave frequencies compared to optical frequencies. This difference becomes even more pronounced under adverse weather conditions such as rain or fog \cite{hoijer2019quantum}.
The atmosphere plays a crucial role in affecting quantum beams. Atmospheric losses in quantum signals can be divided into two categories: static components and dynamic components. To model atmospheric absorption and scattering effects under various conditions and across a wide wavelength range, many atmospheric radiative transfer simulation tools have been developed.

Microwave photons are an effective method for connecting superconducting qubit chips in quantum communication. Although thermal noise is higher at microwave frequencies, signal loss in the atmosphere is low. The antenna impedance depends on differences between environments and can be tuned using nanofabrication techniques. Absorption losses due to superconductivity degradation along long transmission lines between cryogenic and room temperatures are modeled by an infinite series of beam splitters, whose overall effect can be represented by a single effective beam splitter. The effective number of thermal photons introduced into the system by this beam splitter: 

\begin{equation}
    \begin{array}{l}
n_{\text{eff}} = \frac{
\int_0^L dx \, \mu(x) \, n(x) \, e^{-\int_0^x dx' \, \mu(x')}
}{
1 - e^{-\int_0^L dx \, \mu(x)}
}
    \end{array}
    \label{eq:number_of_thermal_photons} 
\end{equation}

This expression is general and can be applied to any situation where the temperature profile is known. Let’s choose a simple but useful profile that will allow us to obtain a closed-form expression. If we assume that a length L can be kept below the critical temperature, then we can make the following choice:

\begin{equation}
    \begin{array}{l}
n(x) = n(T_{\mathrm{in}}) + [n(T_{\mathrm{out}}) - n(T_{\mathrm{in}})] \theta(x - L_0),
    \end{array}
    \label{eq:thermal_photons_simp1} 
\end{equation}
\begin{equation}
    \begin{array}{l}
\mu(x) = \mu_{\mathrm{in}} + (\mu_{\mathrm{out}} - \mu_{\mathrm{in}}) \theta(x - L_0),
    \end{array}
    \label{eq:thermal_photons_simp2} 
\end{equation}
Here, $\mu$in defines the absorption losses at cryogenic temperatures, while $\mu$out represents the material-related absorption losses at room temperature. In this case, the effective number of thermal photons introduced into the system is as follows:

\begin{equation}
    \begin{array}{l}
n_{\mathrm{eff}} = n(T_{\mathrm{in}}) \frac{
e^{-\mu_{\mathrm{out}}(L - L_0)} (1 - e^{-\mu_{\mathrm{in}} L_0})
}{
1 - e^{-\mu_{\mathrm{in}} L_0} e^{-\mu_{\mathrm{out}} (L - L_0)}
}
+ n(T_{\mathrm{out}}) \frac{
1 - e^{-\mu_{\mathrm{out}} (L - L_0)}
}{
1 - e^{-\mu_{\mathrm{in}} L_0} e^{-\mu_{\mathrm{out}} (L - L_0)}
}
 \end{array}
    \label{eq:effect_number_thermal_photons} 
\end{equation}

Quantum entanglement is a key resource for many quantum information applications, but it is fragile and can easily be destroyed by interactions with the environment, known as quantum decoherence. Entanglement is a highly delicate and unstable state. Therefore, creating and maintaining entangled states is quite challenging. Environmental effects, especially thermal noise, can easily disrupt these states \cite{salmanogli2019modification}. One of the most common issues encountered in quantum radar systems is the propagation of entangled microwave photons through the atmosphere to detect the target. However, since the atmosphere exhibits attenuating effects, the signal generated by the target becomes significantly weakened \cite{salmanogli2020analysis}. Factors such as temperature, pressure, and the interaction of light with solid particles in free space cannot be controlled. Hence, temperature is a particularly critical factor, as thermal photons present in the environment can interact with low-energy entangled photons and disrupt the entanglement. For this reason, entangled photons must be generated at very low temperatures. As the number of thermal photons in the environment increases, entanglement becomes more susceptible to degradation. Consequently, in related studies, the operating temperature is generally kept limited.

After the entangled microwave photons are generated, they are transmitted through the atmosphere to detect the target. To investigate the effects of the atmosphere on these photons, attenuating environments and reflections from the target have been theoretically modeled using beam splitters \cite{salmanogli2020optoelectronic}. Within the framework of quantum mechanics, these effects have been analyzed by considering the atmosphere as a system composed of sequential beam splitters \cite{jeffers1993quantum}. Each beam splitter receives the incoming signal and thermal photons from one side and outputs the desired signal along with environment-induced noise from the other side. The amount of thermal photons present in the environment varies depending on the temperature and altitude of the atmosphere. When entangled photons propagate through the atmosphere, these interactions can significantly weaken the entanglement between the modes.

When examining the scattering effect from the target, the signals propagating through the atmosphere contain both entangled and separable (non-entangled) photons. Depending on atmospheric conditions, the number of entangled photons reaching the target varies. Therefore, the signals interacting with the field generated by the target atom may consist of both entangled and separable photons. This interaction has been modeled using a sequence of beam splitters, where the input corresponds to signals containing thermal photons and the output corresponds to the scattered photons.

The atmospheric environment is considered as a structure containing scattering centers. These centers are represented using beam splitters within a quantum electrodynamics framework \cite{salmanogli2020optoelectronic}. The modeling process begins with the scattering effects in the attenuating medium. For the j-th beam splitter, the inputs are defined as $c_{awj}$,$b_{j}$, and the outputs as $c_{aw(j+1)}$,$a_{sj}$. Therefore, the attenuating medium can be described by a continuous model using a complex coupling coefficient. This model satisfies the condition $|t(\omega)|^2 + |r(\omega)|^2 = 1$ , where $t(\omega)$ represents the transmission coefficient and $r(\omega)$ denotes the reflection coefficient \cite{salmanogli2020optoelectronic}.

\begin{widetext}
\begin{equation}
    \begin{array}{l}
\hat{c}_a(\omega) = e^{[jk_{\mathrm{atm}} - \kappa_{\mathrm{atm}}(\omega)]R} \hat{c}_w(\omega) 
+ j \sqrt{2 \kappa_{\mathrm{atm}}(\omega)} \int_0^R dz \, e^{[jk_{\mathrm{atm}} + \kappa_{\mathrm{atm}}(\omega)](R - z)} \hat{b}(\omega, z)
    \end{array}
    \label{eq:atten_output_operator} 
\end{equation}
\end{widetext}

\begin{figure*}[b]
    \centering
    \includegraphics[width=0.75\textwidth]{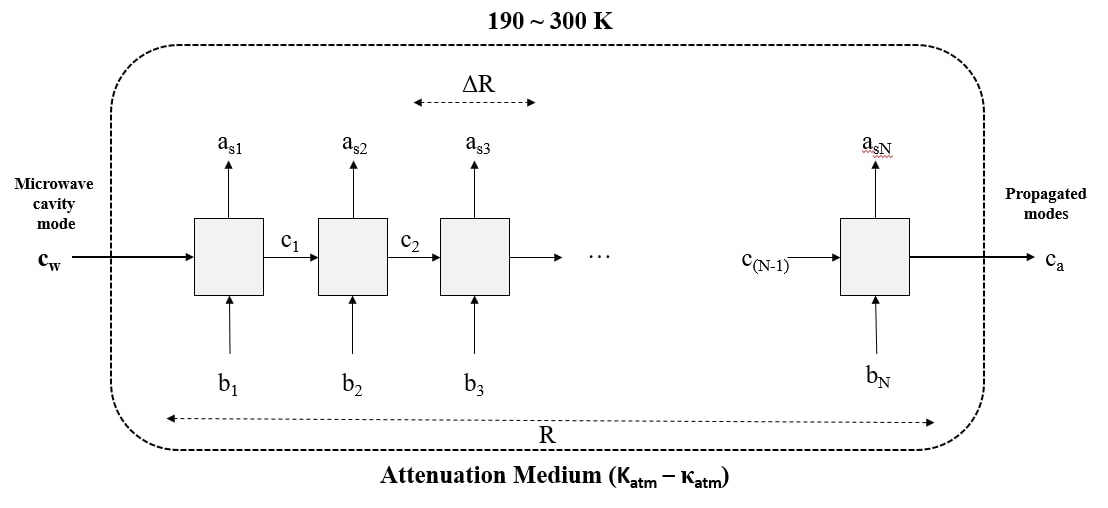}
    \caption{\label{fig:attenuation_medium} Modeling the impact of the atmospheric attenuation medium on microwave cavity modes through beam splitter techniques \cite{Salmanogli2020July}.}
\end{figure*}

$c_a(\omega)$,$K_{\text{atm}}$, and $\kappa_{\text{atm}}(\omega)$ represent the attenuation medium output mode operator, the real component of the wave vector, and the imaginary component of the wave vector, respectively. As illustrated in Fig.\ref{fig:attenuation_medium} , the impact of the atmospheric attenuation medium on microwave cavity modes is modeled using beam splitter techniques.

\begin{figure*}
    \centering
    \includegraphics[width=0.75\textwidth]{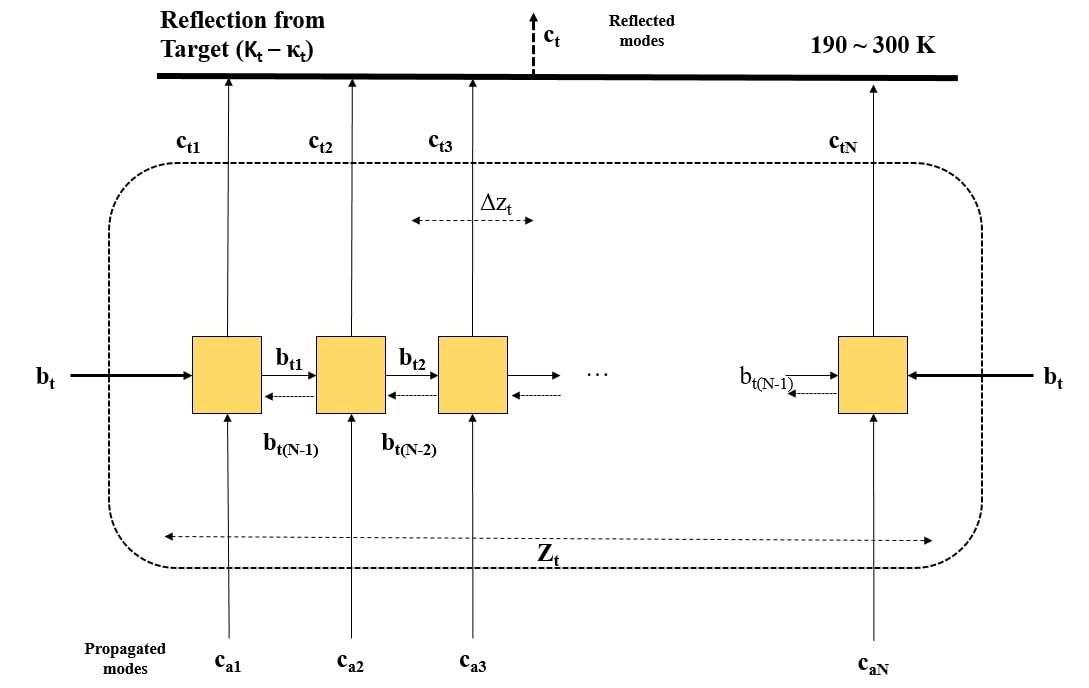}
    \caption{\label{fig:reflection_from_target} Target scattering modeled through beam splitter techniques \cite{Salmanogli2020July}.}
\end{figure*}

Another significant factor in quantum radar is the reflection of the signal from the target. This effect can disrupt entangled states. Similar to the approach used for the attenuating medium, quantum electrodynamics theory is employed here to model the scattering effect. In the quantum framework, reflection from the target is defined as scattering occurring from the target atoms. This process represents the interaction between the electromagnetic quantum field and the quantum fields of the atoms in the target. Thermal photons play a particularly important role during scattering, exerting substantial influence on entanglement. To model this process, the j-th beam splitter (BS) is considered as a reflective element. In this case, the relationship between the input and output is continuously defined as follows \cite{salmanogli2020optoelectronic}:

\begin{widetext}
\begin{equation}
\begin{aligned}
\hat{c}_t(\omega) &= \left[ t(\omega) e^{iK_t\Delta z_t} + 2\kappa_t(\omega)\sqrt{\Delta z_t} \int_0^{z_t} dz \, \big\{ t(\omega) e^{[iK_t - \kappa_t(\omega)]} \big\}^{(z_t - z)} \right] \hat{c}_a(\omega) \\
&\quad + i \sqrt{\kappa_t(\omega)\Delta z_t} \, e^{[iK_t - \kappa_t(\omega)] z_t} \hat{b}_t(\omega)
\end{aligned}
\label{eq:relationship_bw_in_and_out}
\end{equation}
\end{widetext}

where $c_t(\omega)$, $K_t$, and $\kappa_t(\omega)$ denote the target’s scattering output mode operator, the real part of the wave vector, and the imaginary part of the wave vector, respectively. The first term in Eq.\ref{eq:relationship_bw_in_and_out} represents the effect of the imaginary part of the target's dielectric constant. The second term reflects the influence of thermally excited photons. In other words, Eq.\ref{eq:relationship_bw_in_and_out} indicates a significant reduction in the amplitude of the incoming wave, denoted as $c_a(\omega)$. More importantly, the phase information caused by the thermal photons in the second term of Eq.\ref{eq:relationship_bw_in_and_out} can severely disrupt the entanglement. Fig.\ref{fig:reflection_from_target} illustrates the modeling of target scattering using beam splitter techniques.

\begin{figure*}[b]
    \centering
    \includegraphics[width=0.75\textwidth]{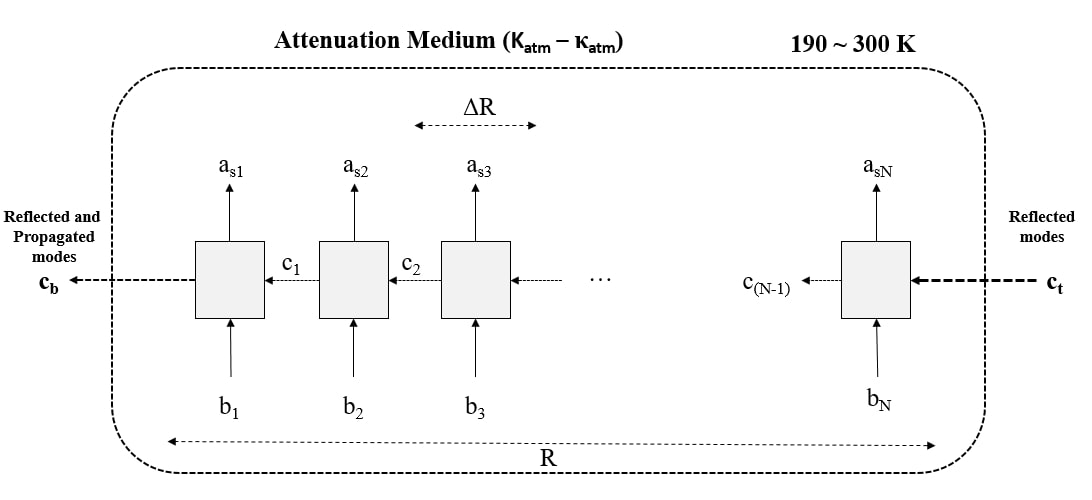}
    \caption{\label{fig:reflected_and_propagated} Representation of scattering signal transmission through the atmosphere as an attenuation medium via beam splitters\cite{Salmanogli2020July}.}
\end{figure*}

Photons reflected from the target propagate again through the atmosphere. Therefore, to complete the quantum radar process, the effect of the atmosphere must be considered once more. However, there is a significant difference here: while the input signal initially entering the atmosphere contains a substantial number of entangled photons, the photons reflected from the target typically lose their entanglement. This makes the atmosphere's impact on entanglement even more critical. As depicted in Fig.\ref{fig:reflected_and_propagated}, the transmission of the scattered signal through the atmosphere, treated as an attenuation medium, is represented using beam splitter techniques. Consequently, the system must be designed so that the entanglement is not completely disrupted by external effects, such as atmospheric attenuation and scattering from the target \cite{salmanogli2020optoelectronic}.

The effects of atmospheric and target scattering have been examined using Eq. \ref{eq:atten_output_operator} and Eq. \ref{eq:relationship_bw_in_and_out}. Although these effects are uncontrollable, suitable systems can be designed to generate entangled states at high temperatures. Thus, the harmful impacts of atmospheric and target scattering can be mitigated, preserving entanglement. 

In this study, the effects of the attenuating medium, target scattering, backscattered photons, and atmospheric propagation modes on entanglement were investigated at various distances between the receiver and transmitter. According to Eq. \ref{eq:atten_output_operator} and Eq. \ref{eq:relationship_bw_in_and_out}, as the distance between the transmitter and receiver increases, the number of thermally excited photons increases significantly. This leads to a reduction in the entanglement between the modes.

However, in terms of path loss, because of its proximity to Earth, a low Earth orbit (LEO) experiences minimal diffraction loss, making it particularly useful for quantum key distribution (QKD) experiments. On the other hand, orbits like middle Earth orbit (MEO) and geosynchronous Earth orbit (GEO) have slower speeds that enable maintaining continuous connections for longer periods, supporting extended QKD operations. However, MEO and GEO also present challenges, such as higher radiation levels and increased propagation losses, that need to be considered.
If examined in terms of turbulence, the structure constant $C_n^2$ is a common parameter used to describe wave propagation in random media such as the atmosphere. This constant quantifies fluctuations in the refractive index $n$ caused by irregularities within the medium. To accurately model the atmosphere, it is important to understand how the structure constant varies with altitude and responds to different weather conditions. Performing a precise model of the atmosphere is challenging, and various empirical and parametric models have been developed to describe how $C_n^2$ changes with height. Some commonly used atmospheric models include the Hufnagel-Valley (H-V) model, the Polynomial Regression (PR) model, and the Submarine Laser Communication (SLC) model. Among these, the H-V model is the most widely adopted.

Due to atmospheric turbulence, the beam also experiences increased divergence. To manage this, an aperture size of 15 cm has been chosen as optimal. Smaller apertures result in higher transmission losses, while larger apertures are more susceptible to turbulence, leading to a lower Strehl ratio at the detector. To further enhance performance, additional technologies such as Adaptive Optics (AO) can be implemented. Another major source of loss is atmospheric absorption. This issue has previously been addressed in discussions on quantum interferometric radar. Although adaptive optics can also help mitigate this problem, their use is similarly subject to practical limitations and constraints \cite{torrome2020introduction}.

Studies in this direction can be summarized as follows.

In certain system configurations, the impact of thermal noise on the antenna can be significantly reduced by incorporating cryogenic losses prior to the antenna input \cite{gonzalez2022coplanar,sanz2018challenges}. Instead of assuming that thermal photons enter directly through the antenna, some are absorbed within the cryostat, effectively lowering the thermal noise contribution. Including these losses creates a trade-off between the entanglement degradation introduced by the effective beam splitter and the improved antenna performance due to a reduced thermal photon population. This balance is highly sensitive to the temperature profile along the transmission line, which must be accurately characterized to inform subsequent impedance optimization and modeling of entanglement loss beyond the antenna \cite{gonzalez2022coplanar,sanz2018challenges}.

Recent studies have demonstrated that entanglement-based quantum communication protocols can enable secure transmission even under high-loss and noisy conditions. For instance, the first experimental implementation of a protocol that maintains secure communication despite channel noise exceeding the entanglement-breaking threshold by 8.3 dB was reported in \cite{zhang2013entanglement}. This protocol relies on interference, requiring precise timing between the signal and idler paths, and employs dispersion- compensating fiber to maintain performance. Despite a total measured channel loss of approximately 16.4 dB and an idler loss of 4.1 dB, the system successfully sustained secure communication. These results highlight the practical feasibility of entanglement-based quantum communication in realistic, lossy, and noisy fiber-optic environments \cite{zhang2013entanglement}.

Nonclassical states are fundamental to optical quantum information processing, but their fragility limits their practical use since loss and noise inevitably degrade or destroy their quantum properties. In quantum metrology, nonclassical states offer sensitivity advantages over all classical systems using the same probe energy. However, these improvements are almost always severely reduced by quantum decoherence. In \cite{zhang2015entanglement}, an entanglement-enhanced sensing system that remains robust against decoherence is experimentally demonstrated. Despite facing 14 dB loss and a noise background 75 dB stronger than the returned signal, conditions that typically destroy entanglement, a 20 percent improvement in SNR compared to the best classical system has been achieved. This result suggests that quantum sensing technologies can offer practical advantages under noisy and lossy conditions \cite{zhang2015entanglement}. The performance advantage of QI over classical illumination (CI) with the same probe energy arises from the strong correlations between the returned and stored beams created by the initial entanglement, even when the environment destroys the entanglement itself. Joint measurement leverages this strong quantum signature to outperform classical systems. Here, missing experimental evidence that QI can significantly improve target detection performance compared to the best classical system under conditions that destroy entanglement is provided \cite{zhang2015entanglement}. Measured SNR values for QI closely match theoretical predictions, while classical illumination SNR deviates slightly at the lowest and highest transmission values due to polarization drifts and laser phase instability, respectively. The QI system's SNR advantage holds despite the channel completely destroying the initial signal-idler entanglement, indicating that the initial quantum correlations still provide measurable benefits. This shows that entanglement-based protocols can improve sensing performance under harsh conditions where traditional thinking expects quantum advantages to vanish \cite{zhang2015entanglement}. In summary, the authors implement an entanglement-based sensing protocol that enhances SNR over optimal classical systems even in environments with high loss and strong noise that fully eliminate initial entanglement. Achieving this quantum advantage without compensating for device imperfections challenges the conventional view that quantum benefits disappear under such conditions and motivates further exploration of quantum information techniques for practical applications \cite{zhang2015entanglement}.

Satellite-based quantum communication, particularly quantum key distribution (QKD), is considered a key enabler for global quantum networks\cite{behera2024estimating}. To assess the feasibility of such systems, realistic atmospheric simulations are realized—especially for uplink scenarios where the quantum signal interacts with the atmosphere early in its propagation \cite{behera2024estimating}. A study in \cite{behera2024estimating} validated its simulation methodology using experimental data from the Canary Islands and Canada, and applied it to evaluate three ground stations in India. Through atmospheric modeling and link budget analysis, the study identified IAO Hanle as the most suitable site for uplink-based quantum communication among the considered locations \cite{behera2024estimating}.

In the absence of atmospheric turbulence, the received power at the satellite detector is mainly affected by free-space path loss and optical absorption. Taking into account optical losses in both the transmitter and receiver components, as well as atmospheric absorption, the received power can be calculated accordingly \cite{behera2024estimating}.

\subsection{Quantum Receiver}

In microwave quantum illumination, the quantum receiver is central to extracting performance gains over classical radar systems. The ideal quantum receiver performs a joint measurement between the returned signal and the retained idler mode, leveraging quantum correlations, such as entanglement or squeezing, to discriminate the presence or absence of a target in high noise \cite{guha2009gaussian}. However, realizing such optimal joint measurements at microwave frequencies remains technologically difficult due to the absence of practical quantum memories \cite{jeon2025single}, single-photon detectors \cite{wang2025observing}, and loss in cryogenic microwave transmission \cite{kurpiers2017characterizing}.

To overcome these limitations, researchers have developed classically assisted receiver architectures that approximate quantum advantages without requiring ideal quantum hardware. For example, in a proof-of-principle experiment using the JPC, signal and idler beams are downconverted, digitized, and then post-processed classically to evaluate their second-order quadrature correlations \cite{barzanjeh2020microwave}. This so-called digital receiver allows partial recovery of the quantum illumination benefit, even though it does not implement an entanglement-preserving joint detection. The trade-off is a suboptimal but implementable receiver that can still outperform the classical radar in terms of error probability or receiver operating characteristics.

In practice, the receiver chain integrates a hierarchy of amplifiers in multiple temperature stages. Mostly, started by high-electron-mobility transistors (HEMTs) at the 4 K stage and low-noise amplifiers (LNAs) at room temperature. This sequential chain guarantees the preservation of delicate quantum correlations encoded in the weak microwave signals throughout the amplification process, enabling accurate digitization. The reflected signal, collected by a secondary antenna, is downconverted through an IQ mixer driven by a local oscillator, producing in-phase (I) and quadrature (Q) components. These signals are digitized and stored for correlation analysis with the retained idler, enabling quantum-enhanced detection without the need for optical quantum memories. This hybrid design, which combines cryogenic amplification and classical digital processing, makes the quantum receiver both scalable and robust to loss, thereby bridging theoretical proposals with experimental feasibility in quantum radar.

Other strategies, such as quantum-enhanced noise radar, use TMSV states generated by JPAs, but forego direct joint measurement \cite{chang2019quantum}. Instead, they rely on statistical analysis of the correlations between noisy return signals and stored idler beams, again evaluated digitally.

\subsection{Quantum Signal Processing and FPGA} 

Quantum signal processing in microwave quantum radar systems is fundamentally based on analyzing the statistical structure of quadrature measurements from the \textit{signal} and \textit{idler} fields generated via the TMSV states. These states are modeled using Gaussian quantum optics, which allows complete characterization via covariance matrices of the field quadratures \cite{luong2020quantum}.

The quadrature operators for the signal (\( \hat{a}_s \)) and idler (\( \hat{a}_i \)) modes are defined as:

\begin{equation}
    \hat{x}_{k} = \frac{1}{\sqrt{2}} \left( \hat{a}_{k} + \hat{a}_{k}^\dagger \right), \quad
    \hat{p}_{k} = \frac{-j}{\sqrt{2}} \left( \hat{a}_{k} - \hat{a}_{k}^\dagger \right)
\end{equation}
where \( k = s, i \) denotes the signal and idler, respectively.

The 4×4 covariance matrix \( \mathbf{V} \) for the joint Gaussian state of signal and idler is defined by:

\begin{equation}
    V_{mn} = \frac{1}{2} \left\langle \{ \hat{\xi}_m, \hat{\xi}_n \} \right\rangle - \langle \hat{\xi}_m \rangle \langle \hat{\xi}_n \rangle
\end{equation}
with the quadrature vector:

\begin{equation}
    \hat{\xi} = (\hat{x}_s, \hat{p}_s, \hat{x}_i, \hat{p}_i)^T
\end{equation}

For an ideal TMSV state with squeezing parameter \( r \), the covariance matrix takes the standard form:

\begin{equation}
    \mathbf{V}_{\text{TMSV}} =
\begin{bmatrix}
    \cosh(2r) \, \mathbb{I}_2 & \sinh(2r) \, \sigma_z \\
    \sinh(2r) \, \sigma_z & \cosh(2r) \, \mathbb{I}_2
\end{bmatrix}
\end{equation}
where \( \sigma_z = \text{diag}(1, -1) \) and \( \mathbb{I}_2 \) is the 2×2 identity matrix \cite{luong2023quantum}.

In a quantum radar receiver, these quadratures are sampled using heterodyne or homodyne detection. After the signal interacts with the environment and reflects back, it is combined with the stored idler (or a digitally processed replica) to compute second-order moments and perform detection.

The quantum correlation coefficient \( \rho \) between the signal and idler is often extracted from the off-diagonal terms of the covariance matrix:
\begin{equation}
    \rho = \frac{\text{Cov}(x_s, x_i)}{\sqrt{\text{Var}(x_s) \, \text{Var}(x_i)}}
\end{equation}

This coefficient directly influences the Receiver Operating Characteristic (ROC) curves and the detection performance \cite{luong2020quantum, luong2023quantum}.

The quantum receiver architecture replaces the ideal joint quantum measurement with a classical post-processed measurement scheme, enabled by high-speed analog-to-digital converters (ADC) \cite{barzanjeh2020microwave}. Both the retained idler and the reflected signal are amplified and downconverted to an intermediate frequency, then sampled by high-resolution ADCs operating at 1 GS/s (giga-sample per second). These ADCs capture the in-phase (I) and quadrature (Q) components of each field with sufficient temporal resolution to retain the underlying quantum correlations \cite{barzanjeh2020microwave}. These ADCs capture the I and Q components of each field with sufficient temporal resolution to retain the underlying quantum correlations. The sampled I and Q data are first transferred to a field-programmable gate array (FPGA) embedded on the digitizer board, where initial data buffering and formatting are performed. Subsequently, the data is sent to a host PC for further digital post-processing. 

In the host PC, the digitized I/Q data streams are processed using a correlation-based digital receiver implemented in software \cite{da2010schemes, eichler2011observation, barzanjeh2019stationary, barzanjeh2020microwave}. The receiver computes second-order moments (e.g., covariance between signal and idler quadratures) to estimate the presence or absence of a target. This digital postprocessing step substitutes for a true quantum joint measurement, enabling a practical and scalable realization of quantum illumination in the microwave domain.

\section{Microwave Quantum Radar Experimental Systems} \label{sec4}

\begin{figure*}
    \centering
    \includegraphics[width=0.75\textwidth]{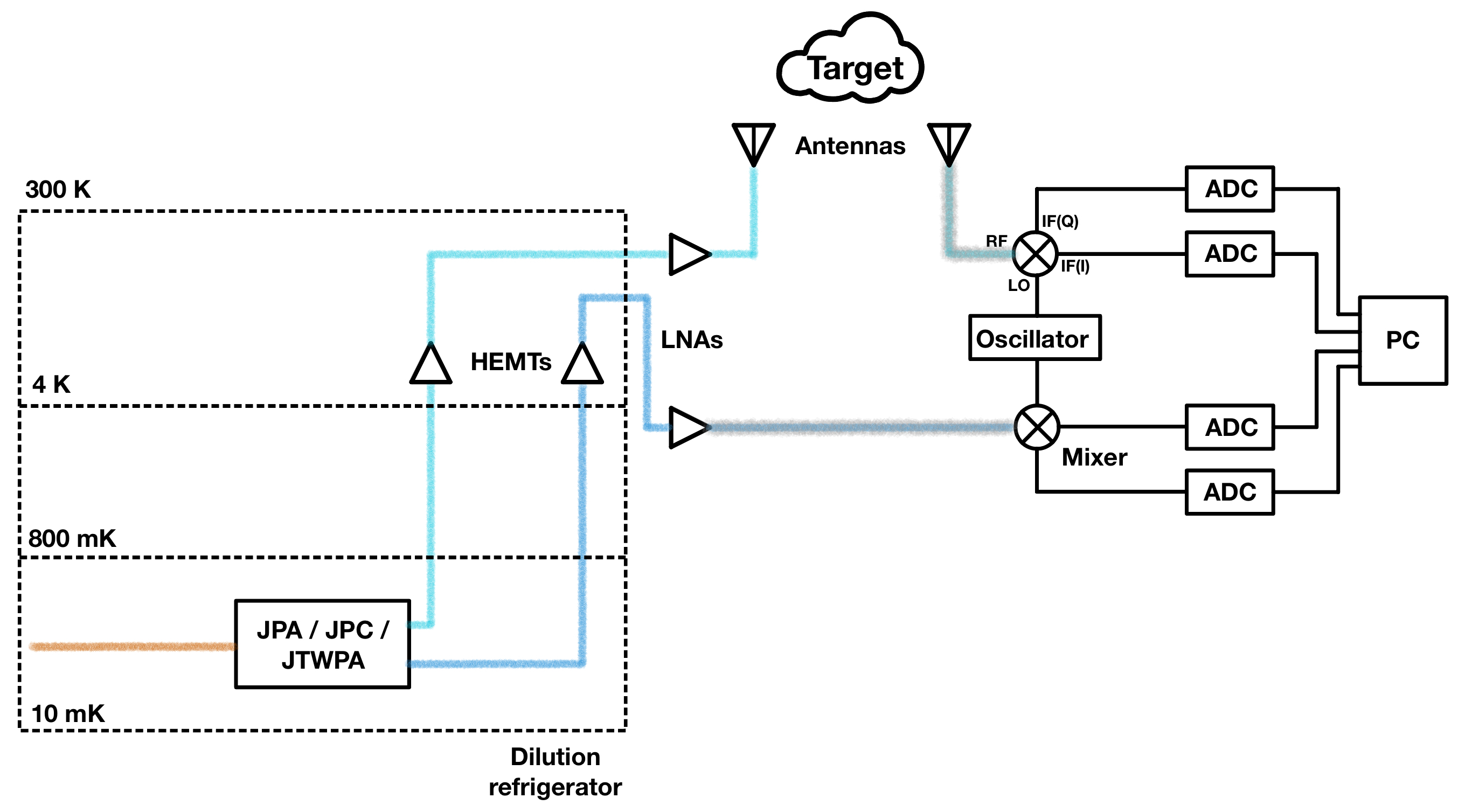}
    \caption{\label{fig:q_experimental} Experimental setup.}
\end{figure*}

In microwave quantum radar systems, achieving reliable detection of weak signals requires high-sensitivity and low-noise amplification. In this context, various amplifier technologies operating at different temperature stages are combined to amplify and preserve quantum-correlated signals. Particularly, HEMT amplifiers operating at the 4 K temperature stage play a critical role in practical implementations. Below, a typical amplification and signal processing architecture used in microwave quantum radar experiments is detailed, with an emphasis on the role of HEMT amplifiers within the system. Figure \ref{fig:q_experimental} illustrates a typical cryogenic measurement and signal acquisition architecture for microwave quantum radar experiments \cite{luong2019receiver, barzanjeh2020microwave}. The setup integrates quantum-limited amplification, low-noise detection, and high-speed digital signal processing to facilitate quantum-enhanced microwave sensing.

At the heart of the system lies the JPA, JPC, or JTWPA operating at 10 mK, accomodated inside a dilution refrigerator. These superconducting quantum devices generate and amplify entangled microwave signal-idler pairs at quantum-limited noise levels.

The outgoing microwave signal travels through multiple temperature stages, passing through HEMT amplifiers at the 4 K stage and low-noise amplifiers (LNAs) at room temperature (300 K) to boost signal strength while preserving phase-sensitive correlations. The signal is then transmitted through the horn antennas toward a target, where weak reflection is expected.

At the 4 K stage, this amplification is carried out by HEMT amplifiers, which are widely used cryogenic devices in microwave quantum sensing and quantum radar systems. HEMTs have very high carrier mobility and low-noise performance at gigahertz frequencies \cite{cha2020300}. While parametric amplifiers at millikelvin stages provide near-quantum-limited gain, HEMTs deliver the secondary amplification necessary to boost weak microwave signals for further processing at room temperature. Although their noise temperature—typically a few kelvin—is higher than that of superconducting amplifiers, HEMTs remain critical for preserving the fragile information carried by entangled or correlated microwave photons, making them indispensable for practical quantum radar implementations.

The reflected signal is received by a second antenna and downconverted using an IQ mixer driven by a local oscillator (LO), producing two intermediate frequency (IF) channels: in-phase (I) and quadrature (Q). These analog signals are then digitized using high-speed ADCs. The digitized I and Q data are routed to a host PC for digital postprocessing, where signal-idler correlations are extracted using second-order moment analysis to determine target presence.

This setup enables the evaluation of quantum illumination protocols without requiring optical quantum memory, making it scalable and more robust to loss.

\section{Conclusion} \label{sec5}

This review has provided a comprehensive examination of microwave quantum radar, spanning from its foundational theoretical principles to the practical challenges and advances in real-world implementations. We began by discussing quantum illumination protocols and the critical role of entanglement in enhancing detection sensitivity beyond classical limits, even in highly lossy and noisy environments. The generation of entangled microwave photon pairs via superconducting devices such as JPAs, JPCs, and JTWPAs at millikelvin temperatures was highlighted as a key enabling technology.

Subsequent sections detailed the architecture of quantum radar subsystems, including quantum receivers that leverage high-speed analog-to-digital converters and classical post-processing to approximate optimal quantum measurements. The integration of cryogenic amplifiers, such as HEMTs and LNAs, plays a vital role in preserving fragile quantum correlations during signal transmission and reflection.

Furthermore, the review addressed environmental effects such as atmospheric attenuation, turbulence, and target scattering, which pose significant challenges to maintaining entanglement over practical distances. Modeling these effects via beam splitter techniques and incorporating empirical atmospheric models provides critical insights for system design and optimization.
Notably, experimental demonstrations have shown that quantum radar protocols can maintain performance advantages under realistic operational conditions, including high loss and strong thermal noise that completely degrade initial entanglement. These findings underscore that initial quantum correlations, even when destroyed by the environment, can still be exploited via joint detection or correlation-based digital receivers to outperform classical radar systems in signal-to-noise ratio and detection sensitivity.

Looking forward, significant technical challenges remain, including the development of scalable quantum memories, low-noise single-photon detectors in the microwave regime, and integrated optoelectronic interfaces. Nonetheless, the steady progress in superconducting device engineering, cryogenic amplification, and advanced digital signal processing points towards the practical realization of microwave quantum radar systems. Moreover, satellite-based quantum communication experiments and atmospheric channel simulations illustrate the feasibility of global quantum networks that can benefit from quantum radar and sensing technologies.

In conclusion, microwave quantum radar stands as a promising frontier that bridges fundamental quantum physics with applied sensing technologies, offering new paradigms in detection performance, robustness, and applications extending from secure communication to environmental monitoring. Continued interdisciplinary research and technological innovation will be essential to unlock the full potential of quantum radar in both terrestrial and space-based platforms.

\bibliography{aipsamp}% Produces the bibliography via BibTeX.

\end{document}